\documentclass[fleqn,usenatbib]{mnras}

\usepackage[T1]{fontenc}
\usepackage{ae,aecompl}

\usepackage{graphicx}	
\usepackage{amsmath}	
\usepackage{amssymb}	
\usepackage{color}
\usepackage{algorithm}
\usepackage{booktabs}
\usepackage[noend]{algpseudocode}
\usepackage{pifont}
\usepackage{soul}
\usepackage{float}
\usepackage{multibib}
\newcites{Appendix}{Appendix References}
\usepackage{natbib}
\usepackage[flushleft]{threeparttable}
\usepackage{mathrsfs}

\graphicspath{{Figures/}}

\newcommand{\vecB}[1]{\mathrm{{\bmath{\mathit{#1}}}}}
\newcommand{\Exp}[1]{\mathrm{\left\langle #1 \right\rangle}}

\renewcommand{\d}[1]{\ensuremath{\operatorname{d}\!{#1}}}

\DeclareMathOperator\M{\mathcal{M}}
\DeclareMathOperator\Ma{\mathcal{M}_{\text{A}}}

\DeclareMathOperator\kperp{k_{\perp}}
\DeclareMathOperator\kpar{k_{\parallel}}

\usepackage{ulem}
\usepackage{xcolor}

\title[Anisotropic supersonic clouds]{Filaments and striations: anisotropies in observed, supersonic, highly-magnetised turbulent clouds}

\author[J. R. Beattie \& C. Federrath]{
James R. Beattie$^{1}$\thanks{E-mail: beattijr@mso.anu.edu.au} and Christoph Federrath$^{1}$\thanks{E-mail: christoph.federrath@anu.edu.au}
\\
$^{1}$Research School of Astronomy and Astrophysics, Australian National University, Canberra, ACT 2611, Australia\\
}

\date{Accepted XXX. Received YYY; in original form ZZZ}

\pubyear{2019}

\begin{document}
\label{firstpage}
\pagerange{\pageref{firstpage}--\pageref{lastpage}}
\maketitle

\begin{abstract}
Stars form in highly-magnetised, supersonic turbulent molecular clouds. Many of the tools and models that we use to carry out star formation studies rely upon the assumption of cloud isotropy. However, structures like high-density filaments in the presence of magnetic fields, and magnetosonic striations introduce anisotropies into the cloud. In this study we use the two-dimensional (2D) power spectrum to perform a systematic analysis of the anisotropies in the column density for a range of Alfv\'en Mach numbers ($\Ma=0.1$--$10$) and turbulent Mach numbers ($\M=2$--$20$), with 20 high-resolution, three-dimensional (3D) turbulent magnetohydrodynamic simulations. We find that for cases with a strong magnetic guide field, corresponding to $\Ma<1$, and $\M\lesssim 4$, the anisotropy in the column density is dominated by thin striations aligned with the magnetic field, while for $\M\gtrsim 4$ the anisotropy is significantly changed by high-density filaments that form perpendicular to the magnetic guide field. Indeed, the strength of the magnetic field controls the degree of anisotropy and whether or not any anisotropy is present, but it is the turbulent motions controlled by $\M$ that determine which kind of anisotropy dominates the morphology of a cloud.
\end{abstract}

\begin{keywords}
turbulence -- magnetohydrodynamics -- ISM: clouds -- ISM: kinematics and dynamics -- ISM: magnetic fields
\end{keywords}

\section{Introduction}\label{introduction}
Stars are born inside high-density regions of molecular H$_2$ clouds (MCs) that have undergone collapse. But cloud collapse is not a straightforward process. MCs are anisotropic, fragmented and undergo supersonic, turbulent motions with dynamically relevant magnetic fields \citep{Elmegreen1996,Elmegreen2004,MacLow2004,Lazarian2004,McKee2007,Crutcher2012,Federrath2012,Padoan2014,Hennebelle2019}. The supersonic turbulent motions play a dual role in forming stars by both mixing the H$_2$ gas, which increases the critical density required for collapse and by creating high-density structures where the cloud can cool and form dense protostellar cores \citep{Krumholz2005,Federrath2012,Federrath2015}.  The role of magnetic fields, however, is less understood. Certainly the presence of magnetic fields suppresses the density fluctuations caused by the turbulent motions (e.g. \citealt{Molina2012}) and reduces the star-formation rate by a factor of a few \citep{Padoan2011,Federrath2012,Federrath2015a,Krumholz2019}. However, magnetic fields may also be responsible for overcoming resistance to star formation, for example, by magnetic reconnection, ambipolar diffusion or magnetic breaking. The complex interplay of these physical processes is still not fully understood \citep{Mouschovias2006,Lazarian2012,Hennebelle2019}. Magnetic fields also may play a role in feedback, for example, by driving protostellar jets that inject mass, momentum and energy into the medium, possibly stirring turbulent motions \citep{Frank2014,Gerrard2019,Krumholz2019,Kurwita2019}. They also seem to play a role in the orientation of filamentary structures and striations in the MCs which may not directly influence the star-forming potential of the cloud, but certainly influences the global structure of the cloud through the density PDF \citep{Planck2016a,Planck2016b,Cox2016,Malinen2016,Soler2017,Tritsis2016,Tritsis2018b,Juan2019}. 
High-density filamentary structures house the local conditions necessary to overcome the turbulent motions, and observations find protostars located at junctions (or hubs) between intersecting filaments \citep{Andre2010,Menschikov2010,Schneider2013,Arzoumanian2018,Tokuda2018,Morales2019,Roy2019,Xu2019}. Indeed, high-density filaments may even play a role in setting the stellar initial mass function \citep{Andre2019}. Striations, in contrast to high-density filaments, most likely form through fast magnetosonic waves, largely independent of the turbulent properties of the cloud \mbox{\citep{Tritsis2016,Tritsis2018b}}, consisting of thin channels of compressed gas that run coincident with the magnetic field direction  \citep{Cox2016,Malinen2016,Soler2017}. In this study, we explore how both striations and high-density filaments form and contribute to the anisotropy in the column density.

\subsection{Anisotropy in MHD Turbulence}\label{sec:AnisoTheory}
Magnetic fields locally\footnote{locally in this context means at length scales where the magnetic tension, $([\nabla \cdot \vecB{B}]\vecB{B})/(4\pi)$, can be neglected, and that are comparable to turbulent eddy scales (\citealt{Burkhart2015} and references therein).} impart anisotropic behaviour on the velocity and density structures that are present in the turbulent clouds \citep{Goldreich1995,Cho2000,Cho2003,Kowal2007,Burkhart2014}. This is because the magnetic field does not disappear at any length scale and flows along the field lines do not feel the pressures and tension from the Lorentz force. What this means is that, locally, length scales perpendicular and parallel to the magnetic field lines become dynamically dissimilar in the cloud. Pioneering work from \cite{Goldreich1995} showed that the anisotropy in the energy spectrum for an incompressible, turbulent magnetohydrodynamical (MHD) fluid with equal energy contributions from the magnetic and turbulent forces is,
\begin{equation}\label{eq:GSscaling}
    \frac{\kpar}{\kperp} \sim \left(L \kperp \right)^{-1/3},
\end{equation}
where $\kpar$ and $\kperp$ are $k$-space scales parallel and perpendicular, respectively, to the mean, local magnetic field, and $L$ is the cloud scale. The anisotropy is introduced from a critical balance between the turnover time of turbulent eddies perpendicular to the magnetic field, $T_{\perp} \sim \ell_{\perp} / V_{\ell}$, where $V_{\ell} \sim \ell_{\perp}^{1/3}$ is the velocity scaling for eddies perpendicular to the field-lines, and the period of the Alfv\'en waves travelling along the field-lines, $T_{\parallel} \sim \ell_{\parallel} / V_{\text{A}}$, where $V_{\text{A}} \sim |\vecB{B}|/\sqrt{\rho}$ is the Alfv\'en velocity, where $|\vecB{B}|$ is the strength of the magnetic field and $\rho$ is the density. Putting this together, 
\begin{equation}
\ell_{\parallel} / V_{\text{A}} \sim \ell_{\perp} / V_{\ell} \sim \ell_{\perp} / \ell_{\perp}^{1/3} \implies \ell_{\parallel} \sim \ell_{\perp}^{2/3}. 
\end{equation}
Furthermore, this means that the anisotropy in the energy spectrum, $\kpar \sim \kperp^{2/3}$, is a function of scale. The anisotropy grows slowly with $\kperp$, i.e. for small length scales in the turbulence we expect stronger anisotropies. This results in the turbulent eddies becoming stretched along the magnetic field and small-scale eddies are stretched more than large-scale eddies. The \cite{Goldreich1995} model has been numerically validated in the incompressible regime \citep{Cho2000,Maron2001,Cho2002} and extended for compressible turbulence in the sub-Alfv\'enic regime $(\Ma < 1$, where $\Ma = (c_s \M) / V_A$ is the Alfv\'en Mach number, $\M = V_L / c_s$ is the sonic Mach number, where $c_s$ is the sound speed in the cloud and $V_L$ is the velocity at cloud scale $L$), and in the super-Alfv\'enic regime $(\Ma > 1$; \citealt{Lazarian1999,Cho2003}). However, the scaling laws predicted in the \cite{Goldreich1995} model are still debated (see discussion in \citealt{Perez2007}, and spectral analysis in \citealt{Boldyrev2006}, for example) and will only be resolved when high-resolution simulations can separate the relevant scaling ranges without ambiguity. 

The anisotropic behaviour of MHD turbulence is of great interest (e.g. \citealt{Bigot2008,Esquivel2011,Burkhart2014,Verdini2015,Tritsis2018b,Hennebelle2019,Xu2019}). It is not only one of the key differences between MHD turbulence and the \cite{Kolmogorov1941} hydrodynamical (HD) turbulence, and leads to many interesting astrophysical phenomena, but also because the ansiotropy can be used as a tool for measuring cloud kinematics and may end up playing an important role in the creation of filaments \citep{Hennebelle2019,Xu2019}. For example, properties of anisotropic density structures, striations, have been used to measure magnetic field strength in the cloud \citep{Tritsis2016,Tritsis2018b}, the velocity anisotropies (centroids in PPV space) have been used to determine the Alfv\'enic regime that the turbulence is in, first by \cite{Esquivel2011} and then extended by \cite{Burkhart2014}. \cite{Xu2019} provides a model for low-density filament orientation caused by anisotropies in the density field of subsonic turbulence. In our study we further explore the anisotropic density structures induced by strong magnetic fields and strong turbulence, systematically across a large parameter range by comparing the three-dimensional (3D) simulation data with line-of-sight projections.

\subsection{Line-of-sight Projections}

Line-of-sight (LOS) position-position (PP) projection data, i.e. the column density, $\Sigma$, can be obtained through molecular lines, dust emission, or dust extinction observations of the MCs \citep{Schneider2015}. The intrinsically 3D density structures appear significantly altered in LOS projections. For example both high and low densities are truncated in the LOS projection (i.e. the variance of the density distribution decreases; \citealt{Brunt2010b,Federrath2013b}) and density variations on large scales are smoothed less than on small scales \citep{Beattie2019a}. Star formation models, and in general models of turbulence are however constructed and tested upon the 3D density and velocity data. Hence reconstructing 3D kinematic or dynamical information from 2D projections is a highly-nontrivial but important task \citep{Larson1981,Federrath2010,Brunt2010b,Brunt2010a,Beaumont2013,Ginsburg2013,Brunt2014,Kainulainen2014,Tritsis2018,Beattie2019a}. There are a number of analytic models using relations in $k$-space to extract meaningful PPP cloud observables, such as the density dispersion and type of turbulent driving (\citealt{Brunt2010a,Brunt2010b,Brunt2014}; who assumed cloud isotropy) empirical methods used to determine the PPP fractal dimension and turbulent Mach number of the clouds from the statistics of the column density \citep{Sanchez2005,Beattie2019a,Beattie2019b}, and correlations in the (bi)spectrum between PP and PPP data \citep{Burkhart2009}. 


In this study we explore the anisotropic behaviour using the 2D power spectrum of the column density in supersonic, sub-Alfv\'enic and super-Alfv\'enic MHD turbulent regimes with many time-realisations in 20, high-resolution, 3D MHD simulations. Our study is organised into the following sections. In \S \ref{sec:Sims} we discuss the 20 supersonic, turbulent MHD cloud models that we use. In \S \ref{sec:ColDensity} we briefly discuss the structures that are present in the column density maps for each of the simulations. Next, in \S \ref{sec:PS} we define the 2D power spectrum that we use to explore the anisotropic structures in each of the column density maps and show how the different regimes give rise to different types of anisotropic behaviours.  In \S \ref{sec:1DPS} we calculate the $\kperp$ and $\kpar$ cascades from the 2D power spectra. In \S \ref{sec:aniso} we quantify the average anisotropy in the power spectra and reveal the $\M$ and $\Ma$ dependency of the anisotropy. Finally, in \S \ref{sec:conclusion} we summarise the key results of the study.

\begin{table*}
\caption{Main simulation parameters and derived quantities.}
\centering
\begin{tabular}{l|cccr@{}l|r@{}lr@{}lr@{}lr@{}l}
\hline
\hline
& & \multicolumn{2}{c}{Mean $\vecB{B}$-field Components} & \multicolumn{6}{c}{rms Quantities} & \multicolumn{4}{c}{Derived Quantities} \\
\hline
\multicolumn{1}{c}{Simulation} & Grid& $\Ma_0$ & $|\vecB{B}_0$| & \multicolumn{2}{c}{$\M$} & \multicolumn{2}{c}{$\Ma$} & \multicolumn{2}{c}{$|\vecB{B}| \, \text{[$\mu$G]}$} & \multicolumn{2}{c}{$|\delta\vecB{B}| \, \text{[$\mu$G]}$} & \multicolumn{2}{c}{$\Exp{a}_{\kperp}$}   \\
\multicolumn{1}{c}{ID} & Res. & $(\pm1\sigma)$ & $\text{[$\mu$G]}$ & \multicolumn{2}{c}{$(\pm1\sigma)$} &\multicolumn{2}{c}{$(\pm1\sigma)$} & \multicolumn{2}{c}{$(\pm 1\sigma)$} & \multicolumn{2}{c}{$(\pm 1\sigma)$} & \multicolumn{2}{c}{$(\pm 1\sigma)$} \\
\multicolumn{1}{c}{(1)} & (2) & (3) & (4) & \multicolumn{2}{c}{(5)} & \multicolumn{2}{c}{(6)} & \multicolumn{2}{c}{(7)} & \multicolumn{2}{c}{(8)} & \multicolumn{2}{c}{(9)} \\
\hline \\ 
\texttt{M2Ma10}\dotfill    & $512^3$   & 9.0 $\pm$ 0.8     & 0.71     & 1.80\,   & $ \pm$ 0.08     & 2.4\, &$ \pm$ 0.2          & 2.6\,     &$ \pm$ 0.1         & 2.0\,  &$ \pm$ 0.1    & 1.03\,     &$ \pm$  0.11 \\[1em]
\texttt{M2Ma2}\dotfill     & $512^3$   & 1.7 $\pm$ 0.1     & 3.54     & 1.66\,   & $ \pm$ 0.05     & 1.32\, &$ \pm$ 0.07        & 4.5\,     &$ \pm$ 0.1         & 1.0\,  &$ \pm$ 0.1    & 0.88\,     &$ \pm$ 0.02  \\[1em]
\texttt{M2Ma1}\dotfill     & $512^3$   & 0.98 $\pm$ 0.07   & 7.09     & 2.0\,    & $ \pm$ 0.1      & 0.94\, &$ \pm$ 0.07        & 7.36\,    &$ \pm$ 0.02        & 0.27\,  &$ \pm$ 0.02  & 0.61\,     &$ \pm$ 0.06  \\[1em]
\texttt{M2Ma0.5}\dotfill   & $512^3$   & 0.54 $\pm$ 0.04   & 14.18    & 2.2\,    & $ \pm$ 0.2      & 0.54\, &$ \pm$ 0.04        & 14.26\,   &$ \pm$ 0.01        & 0.08\, &$ \pm$ 0.01    & 0.5 \,     &$ \pm$ 0.1    \\[1em]
\texttt{M2Ma0.1}\dotfill   & $512^3$   & 0.133 $\pm$ 0.008 & 70.90    & 2.6\,    & $ \pm$ 0.2      & 0.131\, &$ \pm$ 0.01       & 70.902\,  &$ \pm$ 0.001       & 0.002\,  &$ \pm$ 0.001    & 0.57\,     &$ \pm$ 0.07  \\[1em]
\texttt{M4Ma10}\dotfill    & $512^3$   & 9.2 $\pm$ 0.6     & 1.42     & 3.7\,    & $ \pm$ 0.1      &  2.8\, &$ \pm$ 0.2         & 4.6\,    &$ \pm$ 0.1          & 3.2\, &$ \pm$ 0.1    & 1.00\,     &$ \pm$ 0.03  \\[1em]
\texttt{M4Ma2}\dotfill     & $512^3$   & 1.73 $\pm$ 0.07   & 7.09     & 3.5\,    & $ \pm$ 0.1      & 1.43\, &$ \pm$ 0.06        & 8.6\,    &$ \pm$ 0.1          & 1.5\, &$ \pm$ 0.1    & 0.94\,     &$ \pm$ 0.02  \\[1em]
\texttt{M4Ma1}\dotfill     & $512^3$   & 0.95 $\pm$ 0.08   & 14.18    & 3.8\,    & $ \pm$ 0.3      & 0.98\, &$ \pm$ 0.07        & 14.61\,    &$ \pm$ 0.06       & 0.43\, &$ \pm$ 0.06    & 0.80\,     &$ \pm$ 0.08  \\[1em]
\texttt{M4Ma0.5}\dotfill   & $512^3$   & 0.54 $\pm$ 0.03   & 28.36    & 4.4\,    & $ \pm$ 0.2      &  0.54\, &$ \pm$ 0.03       & 28.49\, &$ \pm$ 0.03          & 0.13\, &$ \pm$ 0.03    & 0.7 \,     &$ \pm$ 0.1    \\[1em]
\texttt{M4Ma0.1}\dotfill   & $512^3$   & 0.13 $\pm$ 0.01   & 141.80   & 5.2\,    & $ \pm$ 0.4      & 0.13\, &$ \pm$ 0.01        & 141.800\, &$ \pm$ 0.002       & 0.000\,  &$ \pm$ 0.002    & 0.8 \,     &$ \pm$ 0.1    \\[1em]
\texttt{M10Ma10}\dotfill   & $512^3$   & 9.2 $\pm$ 0.7     & 3.54     & 9.2\,    & $ \pm$ 0.4      & 3.1\, &$ \pm$ 0.2          & 10.6\,    &$ \pm$ 0.4         & 7.1\, &$ \pm$ 0.4   & 1.05\,     &$ \pm$ 0.01  \\[1em]
\texttt{M10Ma2}\dotfill    & $512^3$   & 1.8 $\pm$ 0.1     & 17.72    & 9.0\,    & $ \pm$ 0.4      & 1.5\, &$ \pm$ 0.1          & 21.2\,    &$ \pm$ 0.4         & 3.5\,  &$ \pm$ 0.4   & 1.03\,     &$ \pm$ 0.03  \\[1em]
\texttt{M10Ma1}\dotfill    & $512^3$   & 0.93 $\pm$  0.05  & 35.45    & 9.3\,    & $ \pm$ 0.5      & 0.91\, &$ \pm$ 0.04        & 36.3\,    &$ \pm$ 0.1         & 0.9\, &$ \pm$ 0.1    & 1.13 \,     &$ \pm$ 0.09    \\[1em]
\texttt{M10Ma0.5}\dotfill  & $512^3$   & 0.52 $\pm$ 0.02   & 70.90    & 10.5\,   & $ \pm$ 0.4      & 0.52\, &$ \pm$ 0.02        & 71.13\,    &$ \pm$ 0.06       & 0.23\, &$ \pm$ 0.06    & 1.1 \,     &$ \pm$ 0.2    \\[1em]
\texttt{M10Ma0.1}\dotfill  & $512^3$   & 0.125 $\pm$ 0.006 & 354.49   & 12\,     & $ \pm$ 1        & 0.126\, &$ \pm$ 0.006      & 354.500\, &$ \pm$ 0.002       & 0.010\, &$ \pm$ 0.002    & 1.1 \,     &$ \pm$ 0.3    \\[1em]
\texttt{M20Ma10}\dotfill   & $512^3$   & 9.3 $\pm$ 0.8     & 7.09     & 19\,     & $ \pm$ 1        & 3.4\, &$ \pm$ 0.3          & 19.7\,    &$ \pm$ 0.7         & 12.6\, &$ \pm$ 0.7   & 1.03\,     &$ \pm$ 0.01  \\[1em]
\texttt{M20Ma2}\dotfill    & $512^3$   & 1.8 $\pm$ 0.1     & 35.45    & 18\,     & $ \pm$ 1        &  1.58\, &$ \pm$ 0.09       & 41.2\,    &$ \pm$ 0.4         & 5.8\, &$ \pm$ 0.4   & 1.08\,     &$ \pm$ 0.03  \\[1em]
\texttt{M20Ma1}\dotfill    & $512^3$   & 0.93 $\pm$ 0.03   & 70.90    & 19\,     & $ \pm$ 1        & 0.92\, &$ \pm$ 0.03        & 72.5\,    &$ \pm$ 0.3         & 1.6\, &$ \pm$ 0.3    & 1.2\,  &$ \pm$ 0.12    \\[1em]
\texttt{M20Ma0.5}\dotfill  & $512^3$   & 0.53 $\pm$ 0.02   & 141.80   & 21\,     & $ \pm$ 1        & 0.52\, &$ \pm$ 0.02        & 142.2\,    &$ \pm$ 0.1        & 0.4\,  &$ \pm$ 0.1    & 1.2\,  &$ \pm$ 0.2    \\[1em]
\texttt{M20Ma0.1}\dotfill  & $512^3$   & 0.119 $\pm$ 0.003 & 708.98   & 24\,     & $ \pm$ 1        & 0.119\, &$ \pm$ 0.003      & 708.993\, &$ \pm$ 0.003       & 0.013\, &$ \pm$ 0.003    & 1.3\,     &$ \pm$ 0.3    \\[1em]
\hline 
\hline
\end{tabular} \\
\begin{tablenotes}
\item{\textit{\textbf{Notes:}} For each simulation we extract 51 realisations at 0.1$\, T$ intervals, where $T$ is the turbulent turnover time, between $5 \, T$ and $10 \, T$. All $1\sigma$ fluctuations listed are from the time-averaging over $5\, T$. Column (1): the simulation ID. Column (2): the native resolution of the 3D simulation grid. Column (3): the Alfv\'en Mach number for the mean-$\vecB{B}$ component, $\vecB{B}_0$, $\Ma_0 = (2c_s\M\sqrt{\pi \rho_0}) / |\vecB{B}_0|$, where $\rho_0$ is the mean density, $c_s$ is the sound speed and $\M$ is the turbulent Mach number. Column (4): the strength of the mean magnetic field in direction $\hat{z}$, $\vecB{B}_0 = 2c_s\sqrt{\pi\rho_0}(\M/\Ma_0)\hat{\vecB{z}}$. Column (5): the rms turbulent Mach number. Column (6): the rms Alfv\'en Mach number, $\Ma = (2c_s\M\sqrt{\pi \rho_0}) / |\vecB{B}|$. Column (7): the rms magnetic field strength. Column (8): the turbulent magnetic field component $\delta\vecB{B}$ from the relation $\vecB{B} = \vecB{B}_0 + \delta\vecB{B}$. Column (9): the average anisotropy over all $\kperp $length scales in the zero-mean transformed column density, $\Sigma/\Sigma_0 - 1$, where $\Sigma_0$ is the mean column density. $\kperp$ corresponds to wave vectors perpendicular to the magnetic field direction.}
\end{tablenotes}
\label{tb:simtab}
\end{table*}

\section{Magnetised, Turbulent Molecular Cloud Models} \label{sec:Sims}

\subsection{MHD Model}
In this study we analyse the column densities of 20 high-resolution, 3D turbulent, ideal magnetohydrodynamical (MHD) simulations of quiescent (non-star-forming), supersonic molecular clouds, with no self-gravitation and isothermal equation of state (EOS). A comprehensive parameter set, including mean-field, root-mean-squared (RMS) and derived components for each of the molecular cloud models is listed in Table \ref{tb:simtab}. We use a modified version of \textsc{flash} based on version 4.0.1 \citep{Fryxell2000,Dubey2008} to solve the compressible, ideal MHD equations, 
\begin{align}
    \frac{\partial \rho}{\partial t} + \nabla\cdot(\rho \vecB{v}) &= 0, \\
    \rho\left( \frac{\partial}{\partial t} + \vecB{v}\cdot\nabla \right) \vecB{v} &= \frac{(\vecB{B} \cdot \nabla)\vecB{B}}{4\pi} - \nabla P_{*} + \rho \vecB{F},\label{eq:MHD1} \\ 
    \frac{\partial \vecB{B}}{\partial t} &= \nabla \times (\vecB{v} \times \vecB{B}), \\
    \nabla \cdot \vecB{B} &= 0, \label{eq:MHD3}
\end{align}
where $\vecB{B} = \vecB{B}_0 + \delta\vecB{B}$ is the magnetic field, made from mean-field, $\vecB{B}_0$, and fluctuating, $\delta\vecB{B}$, components, $\rho$ is the density field, $\vecB{v}$ is the velocity field and $P_* = c_s^2\rho + |\vecB{B}|^2/(8\pi)$ is the pressure, which is the sum of the thermal pressure $c_s^2\rho$, where $c_s$ is the sound speed (i.e. the isothermal EOS), which is normalised to $c_s = 1$ in our simulations so all velocities are in units of $\M$, and magnetic pressure, $|\vecB{B}|^2/(8\pi)$. The reader should note that since we have an isothermal EOS we need not include the energy equation in our MHD model, since the isothermal system is completely closed, even in the absence of the energy equation. We solve the MHD equations in a 3D box with dimensions $L \times L \times L$, on a uniform grid with resolution $512^3$, and with periodic boundary conditions. \\

\subsection{Turbulent Driving}\label{sec:TurbDrive}
The $\vecB{F}$ in Equation \ref{eq:MHD1} is an Ornstein-Uhlenbeck (OU) process that satisfies the stochastic differential equation,
\begin{equation} \label{eq:OU}
    \d{\hat{\vecB{F}}}(\vecB{k},t) = F_0(\vecB{k}) \mathscr{P}_{ij} (\vecB{k}) \d{\vecB{W}}(t) - \hat{\vecB{F}}(\vecB{k},t)\frac{\d{t}}{T},
\end{equation}
where $\vecB{W}$(t) is a Wiener process which creates a random Gaussian increment for the forcing field, $\vecB{F}$. The increment is then projected onto the forcing field in $k$-space at $k \sim 2$, i.e. $\sim 1/2$ the box-size, using a projection tensor, $\mathscr{P}_{ij}$,
\begin{equation}
    \mathscr{P}_{ij} = \zeta \left(\delta_{ij} + \frac{k_ik_j}{|k|^2} \right) + (1 - \zeta)\frac{k_ik_j}{|k|^2},
\end{equation}
where $\delta_{ij}$ is the Kronecker delta tensor. We control the contribution from each of the driving modes through the $\zeta$ parameter, e.g. $\zeta = 1$ is for purely solenoidal driving in $\vecB{F}$, and $\zeta = 0$ produces purely compressive driving (see \citealt{Federrath2008,Federrath2009,Federrath2010} for a detailed discussion of the driving). Here we choose a natural mixture of the two modes, $\zeta = 0.5$ \citep{Federrath2010}. The $T$ in Equation \ref{eq:OU} is the autocorrelation timescale of the OU process. For example, if we only consider the second term in Equation~\ref{eq:OU},
\begin{equation}
    \hat{\vecB{F}}(\vecB{k},t) \sim F_0(\vecB{k}) \exp\left\{ - t/ T \right\},
\end{equation}
the autocorrelation timescale defines the characteristic time that the field has lost $1/e$ of its previous structure. The timescale is equal to $T = L /(2c_s \M)$, hence we use $T$ to set the desired turbulent Mach number of the cloud model. For the 20 simulations we vary $\M$ between 2 and 20, encompassing the range of observed $\M$ values for molecular clouds (e.g. \citealt{Schneider2013,Federrath2016b,Orkisz2017}).

\subsection{Initial Conditions and Magnetic Fields}

The initial magnetic field $\vecB{B}$ in Equations \mbox{\ref{eq:MHD1}--\ref{eq:MHD3}} is set to a uniform value with field lines threaded through the $\hat{\vecB{z}}$ direction of the cloud. We set the mean-field component of $\vecB{B}$, i.e., $\vecB{B}_0 = 2c_s\sqrt{\pi\rho_0}(\M / \Ma_0) = B_z$, where $\Ma_0$ is the mean-field Alfv\'en Mach number. Hence by setting $\M$ and $\vecB{B}_0$ we can set the target $\Ma_0$. We explore the supersonic gas dynamics of the clouds in the highly-magnetised, sub-Alfv\'enic regime, $\Ma_0 \approx 0.1$ and 0.5, and in the trans- and super-Alfv\'enic regimes, $\Ma_0 \approx 1$, 2 and 10. These values allow us to explore the transition between the two regimes, which is important to see how the anisotropic behaviour of the column densities changes as the magnetic and thermal energies contribute in different proportions to the cloud dynamics.

We run the simulations to 10$\,T$, where $T$ is eddy the turnover time, equivalent to the autocorrelation timescale discussed in \S\ref{sec:TurbDrive}. We extract column density maps, $\Sigma$, from the volume density along the $\hat{\vecB{y}}$ direction every $0.1 \, T$, between $5 \, T$ and $10 \, T$, ensuring that the turbulent cloud is in a fully-developed, stationary statistical regime \citep{Federrath2009,Price2010}. We however perform the same analysis outlined in the forthcoming sections, \S\ref{sec:PS} and \S\ref{sec:aniso}, on different column density projection angles in Appendix \ref{sec:AppProj}. It also ensures that the initial conditions, $\rho(x,y,z, t=0) = \rho_0 = 1$, where $\rho_0$ is the mean density of the cloud, $\vecB{v}(t=0) = 0$ and $\vecB{B}_0(t=0) = B_z \hat{\vecB{z}}$, do not influence the statistics of the flow that we analyse. Our simulations are dimensionalised so that $\rho(x,y,z)$ is in units of mean density, $\rho_0$. The reason for this is so that our results do not depend upon choice of $\rho_0$ and are solely determined by the $\M$ and $\Ma_0$, i.e. we could scale $\rho_0$ to any value as long as we also scale $|\vecB{B}|$, $c_s$, and $T$, the autocorrelation time, such that $\M$ and $\Ma$ stay the same. We extract column density maps, which are in the same dimensionlised form as, $\rho(x,y,z)$, every $0.1 \, T$, which results in a total of 51 realisations of the column density for each of the simulations. 

\section{Column Density Maps}\label{sec:ColDensity}

\begin{figure*}
    \centering
    \includegraphics[width=\linewidth]{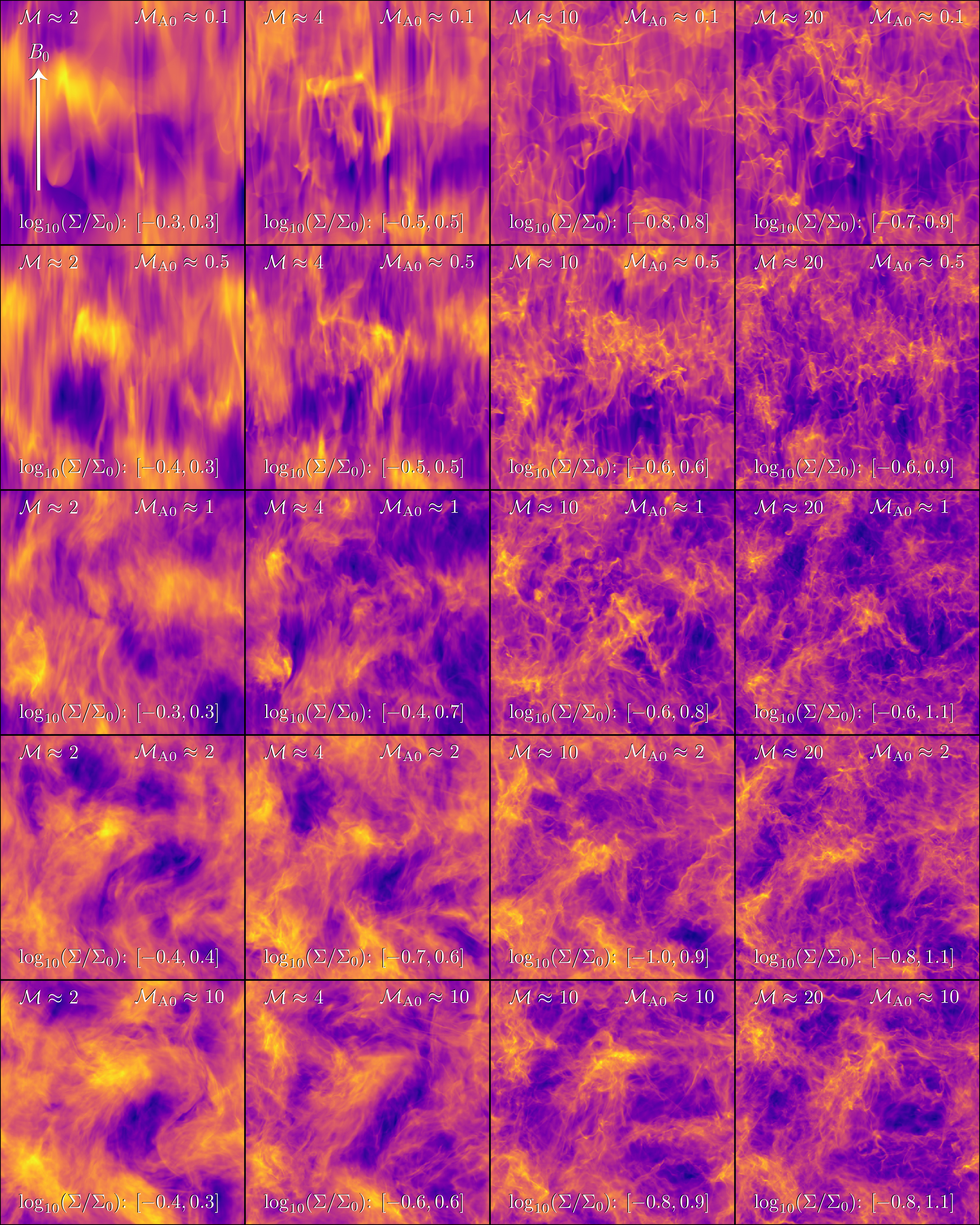}
    \caption{A single realisation at $t/T = 5$ of the $xz$-projected column densities from the 20 simulations analysed in this study. The column density maps are shown on a logarithmic scale and in units of the mean column density, $\log_{10}\left( \Sigma / \Sigma_0 \right)$, with increasing $\M$ (shown in the top-left corner) from left to right and decreasing $\Ma_0$ (shown in the top-right corner) from top to bottom. The magnetic field is oriented along the $z$-axis (up the page). An animation of this figure, revealing the dynamics, is available in the online version.}
    \label{fig:fig1}
\end{figure*}

Figure \ref{fig:fig1} shows a single realisation of the rich structure of the column density at $5\,T$. From left-to-right in the plot, $\M$ increases from 2 to 20, and from top-to-bottom, $\Ma$ increases from 0.1 to 10. Hence the strongest $\vecB{B}$-field and turbulence is in the top-right corner, and the weakest in the bottom-left corner. In the strong $\vecB{B}$-field, low-$\M$ models we find significant striations \emph{parallel} to the magnetic field. By contrast, in the strong $\vecB{B}$-field, high-$\M$ models, high-density filamentary structures and low-density voids develop \emph{perpendicular} to the magnetic field. The parallel striations dominate the morphology of the density field for the $\M = 2$ simulation, while the high-density filamentary structures dominate the column density for the $\M \geq 4$ simulations. This suggests that there is an interplay between $\Ma_0$ and $\M$ on the mode of alignment for filaments with the $\vecB{B}$-field. The strong-field simulations are the most interesting for the purpose of exploring the anisotropic behaviour, whereas the weak-field simulations, i.e. high-$\Ma_0$, qualitatively appear to be in an isotropic regime, where anisotropic structures are present, but they are orientated randomly, contributing to no systematic, preferred direction. To better understand and quantify how the anisotropic structures in the sub-Alfv\'enic regime contribute to the observed properties of the column density we now explore the 2D power spectrum.

\begin{figure*}
    \centering
    \includegraphics[width=\linewidth]{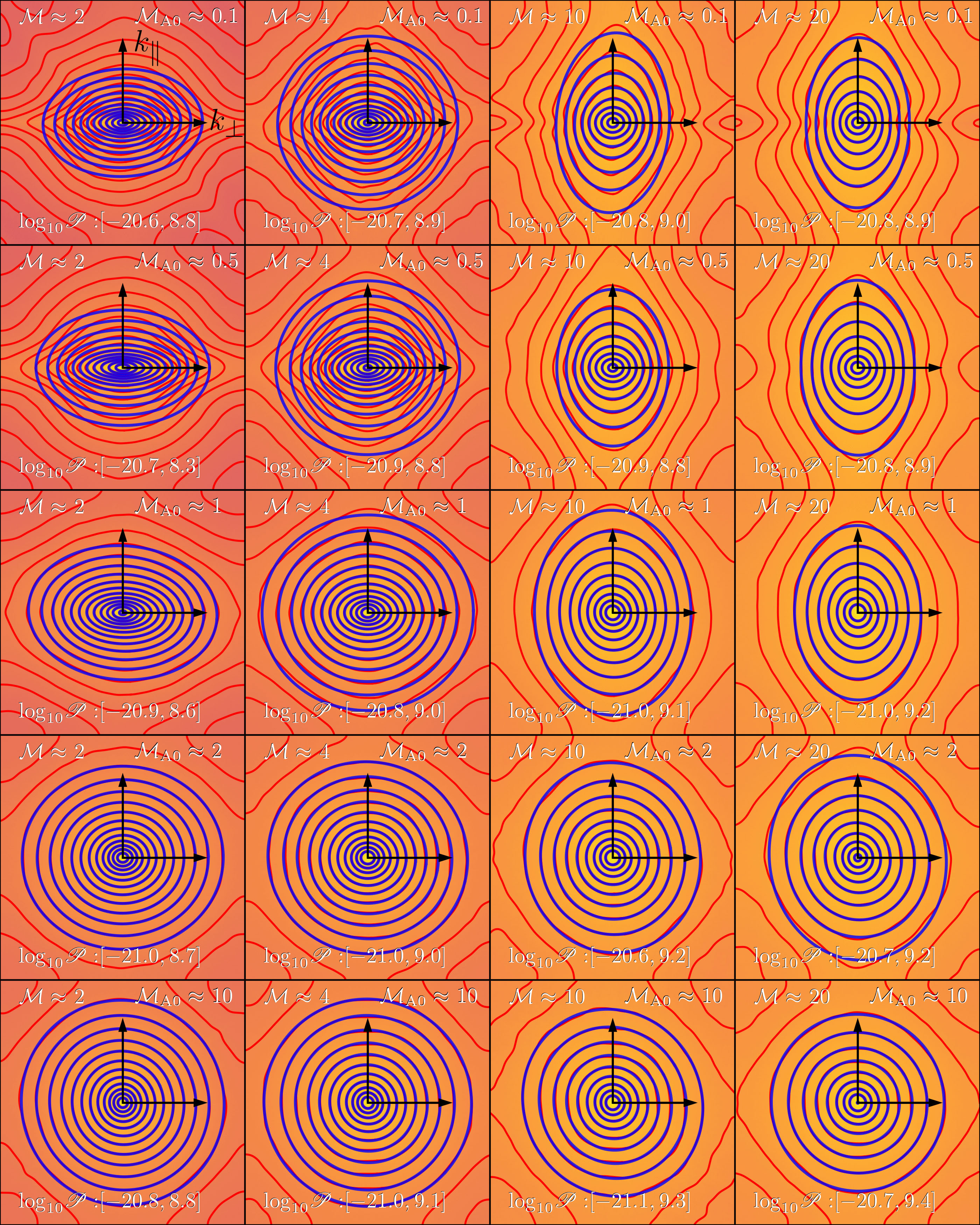}
    \caption{2D power spectrum of the zero-mean column density defined in Equation~(\ref{eq:meanfield}), averaged over 5$\,T$. Each panel shows a different simulation, with the sonic and Alfv\'en Mach number labelled at the top of each panel as in Figure~\ref{fig:fig1}. The colour range is logarithmically scaled from red to yellow, with the range shown at the bottom of each panel. Smoothed iso-surfaces of power are plotted every half order of magnitude, shown with red contours. The $\vecB{k}_\perp, \vecB{k}_\parallel$ origin is shown with the black coordinate system at the centre of each of the power spectra. Ellipses are fit to the iso-surfaces, shown in blue, to approximate the anisotropy parameter discussed in the main text.}
    \label{fig:fig2}
\end{figure*}

\section{The 2D Power Spectra}\label{sec:PS}
The power spectrum allows us to reveal order out of complicated and stochastic structures that are mostly intangible in real space. There are, however, also techniques for directly analysing real-space structures, such as structure functions, topological \citep{Appleby2018,Henderson2019}, and fractal methods \citep{Scalo1990,Elmegreen1996,Stutzki1998,Kowal2007,Federrath2009,Roman-Duval2010,Donovan-Meyer2013,Konstandin2016,Beattie2019b}. The power spectrum has a special role in the study of turbulence, since turbulence models rely heavily upon understanding flow characteristics on different length scales in real space, e.g., on the driving scale, in the inertial (for incompressible flows) scaling range (cascade) of turbulence, and on the dissipation scale \citep{Kolmogorov1941,Burgers1948}. The benefit of using the power spectrum to describe these scales is that we can organise them into $k$-modes, symmetrical (in the isotropic case) about an origin, rather than distributed through real space. However, before constructing the power spectrum we first must consider the quantity that we are taking the power spectra of, i.e. in this study, the column density. 

We transform the column density into a zero-mean field using the transform, 
\begin{equation} \label{eq:meanfield}
\xi(x,y) = \Sigma(x,y)/\Sigma_0 - 1,    
\end{equation}
where $\Sigma_0$ is the mean of the column density. Zero-mean fields are useful for constructing a power spectrum of the field. This is because one can use the power spectrum to measure moments of the field \citep{Brunt2010a,Brunt2010b,Konstandin2016}, e.g., Parsevel's theorem applied to a stochastic field $f$ is, 
\begin{equation}
    \int_{\forall k} \d{k} \, \mathscr{P}_{f} = \Exp{ f^2 },
\end{equation}
where $\Exp{\hdots}$ is the ensemble average operator. We can relate Parsevel's theorem to the 1$^{\text{st}}$ and 2$^{\text{nd}}$ moments of the field by adding and subtracting the square of the 1$^{\text{st}}$ moment, $\Exp{ f }^2$, 
\begin{equation}
    \int_{\forall k} \d{k} \, \mathscr{P}_{f}(k) = \Exp{ f }^2 + \Exp{ f^2 } - \Exp{ f }^2 = \mu^2 + \sigma^2,
\end{equation}
where $\mu$ is the mean (1$^{\text{st}}$  moment) and $\sigma^2$ is the variance. For a zero-mean field, $g$, where $\mu = 0$ 
\begin{equation}
    \int_{\forall k} \d{k} \, \mathscr{P}_{g}(k) = \sigma^2.
\end{equation}
Zero-mean fields are therefore of special interest, since the integral of the power spectrum for a zero-mean field is exactly the variance of the field. \cite{Brunt2010a} and \cite{Federrath2016b} use this property to reconstruct 3D volume-density distributions from the 2D column density distribution (note however that \cite{Brunt2010a} removes the mean in $k$-space rather than in real space). All the power spectra that we construct in this study will be of the zero-mean transformed column density, as defined in Equation~(\ref{eq:meanfield}).

\begin{figure*}
    \centering
    \includegraphics[width=\linewidth]{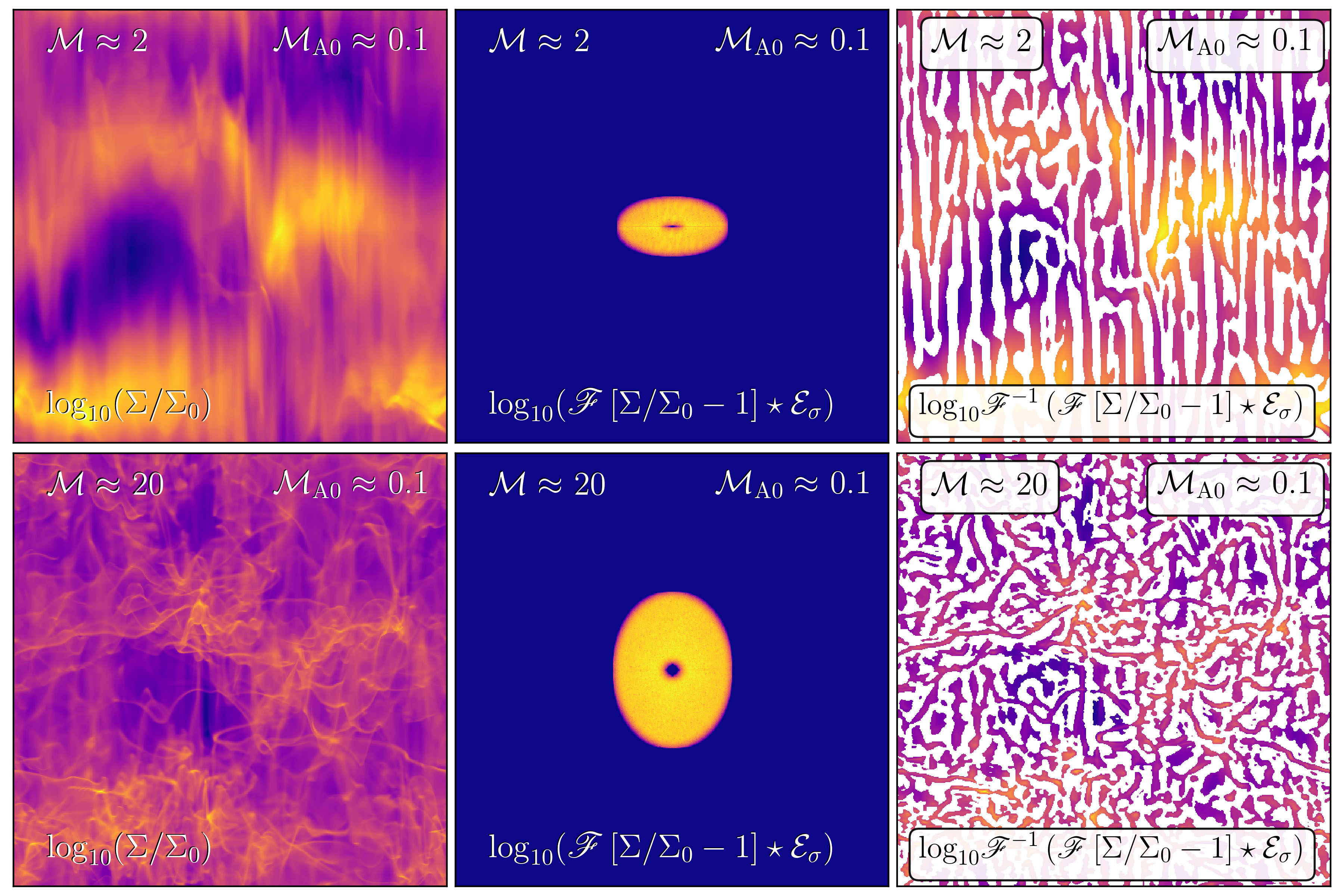}
    \caption{The anisotropic structures that dominate the morphology of the 2D power spectrum for the column density revealed in real space. We show the two sub-Alfv\'enic cases: simulations \texttt{M2Ma0.1} (top panels) and \texttt{M20Ma0.1} (bottom panels). In the first column we show the column densities for each simulation, discussed in \S\ref{sec:ColDensity}. In the second column we show Fourier transforms (Equation \ref{eq:FT}) of the zero-mean transformed column densities (Equation \ref{eq:meanfield}) convolved with the annulus, $\mathcal{E}_{\sigma}$ (discussed in more detail in Appendix \ref{sec:invert}), derived from the elliptic fits shown in Figure \ref{fig:fig2}. In the third column we show the column density structure corresponding to the modes in each of the annulus, i.e., the real-space structures responsible for the anisotropies in the 2D power spectra discussed in \S\ref{sec:PS}. For \texttt{M2Ma0.1} we find highly orientated, thin striations that follow the direction of the $\vecB{B}$-field. By contrast, for \texttt{M20Ma0.1}, we additionally see high-density filaments aligned perpendicular to the $\vecB{B}$-field. An animation of this figure over the full 10$\,T$ is available in the online version.}
    \label{fig:fig3}
\end{figure*}

\subsection{Construction of the Power Spectra}\label{sec:constructPS}
The zero-mean transformed column density, $\xi(x,y)$, is a discrete field with a finite number of resolution elements. Hence we must use the discrete Fourier transform, $\mathscr{F}[\hdots]$, to map the density into $k$-space,
\begin{equation}\label{eq:FT}
    \hat{\xi}(\vecB{k}_{xy}) = \mathscr{F}\left[ \xi(\vecB{\ell}_{xy}) \right] =  \frac{1}{N^2} \displaystyle\sum^{N-1}_{x = 0} \sum^{N-1}_{y = 0} \xi(\vecB{\ell}_{xy}) \exp\left\{ -i\frac{\vecB{\ell}_{xy} \cdot \vecB{k}_{xy} }{N}  \right\}, 
\end{equation}
where $\vecB{\ell}_{xy}$ is the real space vector with coordinates $(x,y)$, $\vecB{k}_{xy}$ is the corresponding $k$-space vector, $\vecB{k}_{xy} = (k_x,k_y)$, $N$ is the grid size in one dimension (i.e., $N = 512$ for all simulations) and $i \equiv \sqrt{-1}$. The 2D column-density power spectrum is the square of the Fourier transform, hence, 
\begin{equation}
    \mathscr{P}_{\xi}(\vecB{k}) = \hat{\xi} \cdot \hat{\xi}^* = \mathscr{P}_{\Sigma/\Sigma_0 -1}, 
\end{equation}
where $\hat{\xi}^*$ is the complex conjugate of $\hat{\xi}$. We show the time-averaged (across all available turnover time realisations) 2D power spectra of the zero-mean column density for each simulation in Figure \ref{fig:fig2}. Shown in red are contours of equal power, plotted every half order of magnitude in power.

For each of the closed contours we fit an ellipse through five degrees of freedom following the generalised eigensystem method outlined in \cite{Fitzgibbon1996}. The five degrees of freedom are translations in $\kpar$ and $\kperp$, rotations of the ellipse in the $\kpar$-$\kperp$-plane, and the modulus of the major and minor principle axes of the ellipse. The purpose of the fit is to approximate the power spectrum with a morphology that captures the main anisotropic features at each $\kperp$ scale. In particular, we want to know how the power spectrum varies along $\kpar$, compared with $\kperp$, at each of the equi-power contours. The elliptic fits lead to a canonical anisotropy parameter, 
\begin{equation} \label{eq:a}
    a \equiv \frac{e_{\parallel}}{e_{\perp}},  
\end{equation}
where $e_{\parallel}$ is the principle axis of the ellipse along the $\kpar$ direction, and $e_{\perp}$ is the principle axis along the $\kperp$ direction. This means that for $a < 1$, $e_{\parallel} < e_{\perp}$, hence the power spectrum is elongated along the $\kperp$ axis. Likewise, $a > 1$ implies $e_{\parallel} > e_{\perp}$, which corresponds to elongation along the $\kpar$ axis. For $a = 1$ (circle), the power spectrum is isotropic. We note that $a$ may depend upon the length scale, i.e., what is isotropic on large length scales need not be isotropic on small length scales or vice versa. This means that $a$ may change for different ellipses, at different $\kperp$ scales in the same power spectra. We show this by choosing to write $a$ as a function of $\kperp$, $a(\kperp)$, consistent with the MHD theory outlined in \S\ref{sec:AnisoTheory}.

In the following sections we discuss the results from the elliptic fits to the power spectra. In \S\ref{sec:superAlf} we focus on the super-Alfv\'enic simulations, and in \S\ref{sec:subAlf} we discuss the sub-Alfv\'enic simulations.

\subsection{Super-Alfv\'enic Regime}\label{sec:superAlf} 
We start our analysis in the bottom eight panels of Figure \ref{fig:fig2}. In these simulations, the magnetic field is weak, placing the turbulence in the super-Alfv\'enic regime ($\Ma_0 > 1$). Our elliptic fits to the contours reveal that on all $k$-scales the power spectrum remains almost entirely circular, i.e., the ratio between $e_{\parallel}\sim e_{\perp}$, and thus, $a \sim 1$. Perfect rotational symmetry of the power spectrum indicates isotropic statistics, reminiscent of hydrodynamical (non-magnetized) turbulence. We can state this mathematically as
\begin{align}\label{eq:isotropic}
    \lim_{\Ma_0 > 1} a(\kperp) \sim 1,
\end{align}
which means that in the super-Alfv\'enic regime, the anisotropy factor $a \sim 1$ for all length scales $\kperp$. This situation changes with increasing magnetic guide field, introducing scale-dependent anisotropies, as discussed in the next subsection.

\subsection{Trans and Sub-Alfv\'enic Regime}\label{sec:subAlf}
Now we move onto the trans ($\Ma_0 \sim 1$) and sub-Alfv\'enic ($\Ma_0 < 1$) turbulent regimes, corresponding to the top 12 panels in Figure \ref{fig:fig2}. On quick inspection one can note two distinct shapes emerging in these regimes. Ellipses stretched along the $\kperp$ axis ($a < 1$) are present in the fits for the $\M \lesssim 4$ simulations, and ellipses stretched along the $\kpar$ axis ($a > 1$) for the $\M \gtrsim 4$ simulations. These are two distinct anisotropies that we will discuss in the next two subsections in detail. We begin with the low-$\M$ simulations.

\subsubsection{$\M \lesssim 4$}\label{sec:less4}
For $\M \lesssim 4$ (top-left corner of Figure \ref{fig:fig2}) the power spectrum is elongated along the $\kperp$ axis. This means that length scales along the magnetic field carry more power on the same length scales perpendicular to the magnetic field, corresponding to an anisotropic factor $a < 1$, as described in \S\ref{sec:superAlf}. We observe the same kind of anisotropy at both large $k$, $\sim k_{\text{max}}$ (small spatial scales) and small $k$, $\sim k_{\text{min}}$ (box-size spatial scales). This is especially obvious in the \texttt{M2Ma0.1} and \texttt{M2Ma0.5} simulations, where the anisotropy grows on smaller $k$ scales (the 2D power spectrum becomes more elongated on large spatial scales, corresponding to small $k$). We state this concisely,
\begin{equation}\label{eq:lowMachAniso}
    a(\kperp \sim k_{\text{min}}) < 1,
\end{equation}
at small $k$, but also at large $k$, hence
\begin{equation}\label{eq:highAniso}
    a(\kperp > k_{\text{min}}) < 1,
\end{equation}
but the anisotropy is strongest at small $k$, hence,
\begin{equation}\label{eq:scaleAniso}
    a(\kperp \sim k_{\text{min}}) < a(\kperp > k_{\text{min}}) < 1.
\end{equation}
Returning to Figure~\ref{fig:fig1} briefly, we can hypothesise that in real space this corresponds to the density striations parallel to the $\vecB{B}$-field dominating the morphology of the column density on $L$ scales (the box-size) and throughout all of the smaller scales local to $\vecB{B}$. Striations are most likely formed through fast magnetosonic waves moving perpendicular to the $\vecB{B}$-field \citep{Tritsis2016,Tritsis2018b}, which is why we see them dominating some of the sub-Alfv\'enic density structures. The reason why these structures lead to stronger anisotropies at small $k$ is because they span coherently across the entire box-length. Figure~\ref{fig:Appendix1} in the Appendix shows the 3D density structures for the \texttt{M1Ma0.1} and \texttt{M20Ma0.1} simulations. In 3D, the striations appear to be sheets of density that are sheared along the field lines and then twisted tightly around them by the turbulent eddies. We will explore these 3D structures in future studies but for now we keep our focus on the column-density structures.

To test our hypothesis we use the elliptic fits (discussed in \S\ref{sec:constructPS}) in $k$-space to extract the real-space structures. The \texttt{M2Ma0.1} simulation is shown in the top row of Figure \ref{fig:fig3}. The first column is the column density map, the same as in Figure~\ref{fig:fig1}. The second column is the real component of the 2D Fourier transform of $\xi$ (Equation~\ref{eq:FT}), the zero-mean transformed column density (Equation~\ref{eq:meanfield}) with a selected annular region, $\mathcal{E}_{\sigma}$ defined by our elliptic fits\footnote{Here $\sigma$ corresponds to the size of a Gaussian smoothing kernel that we apply to the annular region before doing the real-space inversion. We do this to avoid the Gibbs phenomenon at the edges of the annulus, i.e., $\mathcal{E}_{\sigma} = \mathcal{E} * \mathcal{N}(\sigma)$, where $\mathcal{E}$ is the unsmoothed annulus and $\mathcal{N}(\sigma)$ is a Gaussian smoothing function. We pick the smoothing kernel, $\sigma = 2\d{x}$, where $\d{x}$ is the size of a grid cell, for our problem. For more details on this method we refer to Appendix~\ref{sec:invert}.}. We perform the inverse Fourier transform on only the $k$-modes inside of $\mathcal{E}_{\sigma}$, which are shown in the third column of Figure~\ref{fig:fig3}. This figure reveals that the $a < 1$ anisotropy in the power spectrum is indeed caused by magnetosonic striations running parallel to $\vecB{B}$. 

The striations revealed in Figure \ref{fig:fig3} are strongly aligned with the $\vecB{B}$-field, consistent with previous observational analyses \protect{\citep{Cox2016,Malinen2016,Tritsis2018b}}, which indicates a trans- to sub-Alfv\'enic flow regime. Indeed, the striations are such a highly anisotropic feature of the column density and so strongly aligned with the $\vecB{B}$-field that the 2D power spectrum becomes significantly pinched along the $\kperp$ direction. This is evident in the red contours in the top-left four panels in Figure~\ref{fig:fig2}. As $\kperp$ grows, so do the pinched ends of the contours. What this means is that at smaller real space length scales the densities in the cloud are becoming completely dominated by the magnetic field, i.e., the small-scale structures (Equation~\ref{eq:highAniso}) along the $\vecB{B}$-field essentially become uncorrelated with the hydrodynamics. This is because when a power spectrum lacks diagonal modes, the dynamics in each of the directions is not correlated, i.e., the dynamics along the magnetic field is independent of the dynamics perpendicular to the field \citep{Tennekes1972}. 

As the Mach number increases (i.e., as we move right across the panels in Figure~\ref{fig:fig2}), we quickly see the transition and development of a different kind of anisotropy. The transition occurs at $\M \sim 4$ (the second column of Figure~\ref{fig:fig2}), where the power spectrum is full of uncorrelated parallel and perpendicular anisotropic structures resulting in nearly square-shaped contours. The transition occurs at large $k$ first, with small $k$ (large structures), still being significantly aligned with the magnetic field. This is because there are still strongly-aligned, parallel striations on the box-scale, however, gas is becoming more compressed along the field lines, which generates turbulent shocks, perpendicular to the $\vecB{B}$-field. As $\M$ continues to increase we see this new, $a > 1$ anisotropy beginning to dominate the morphology. We now turn our attention to this type of anisotropy.



\subsubsection{$\M \gtrsim 4$}\label{sec:gtrM4}
For $\M \gtrsim 4$ the power contours in Figure~\ref{fig:fig2} form an $a>1$ anisotropy, where more power is in the $\kperp$ direction than on the same length scale in the $\kpar$ direction. This means that our elliptic fits on the column density power spectrum become elongated along the $\kpar$ direction at large $k$, i.e.,
\begin{equation}
    a(\kperp > k_{\text{min}}) > 1.
\end{equation}
However, at small $k$, our elliptic fits are near circular, i.e., on large scales close to $L$, the power spectra return to being isotropic,
\begin{equation} \label{eq:largeliso}
    a(\kperp \sim k_{\text{min}}) \sim 1,
\end{equation}
in contrast with the $\M \lesssim 4$, sub-Alfv\'enic simulations discussed in \S\ref{sec:less4}, where the anisotropic effects of the magnetic field do not disappear on any scale, large or small. Indeed, Equation \ref{eq:largeliso} is true for all high-$\M$ simulations, where the large-scale turbulent energies overcome the preferential orientation imparted by the magnetic fields,
\begin{align}
    \lim_{\M \gg 1} a(\kperp \sim k_{\text{min}}) \sim 1.
\end{align}
However, the most important feature of the $\M \gtrsim 4$ power spectra is the $a > 1$ anisotropy. The elongation along the $\kpar$ axis, at large $k$ indicates that there are structures perpendicular to the magnetic field that do not span the entire box length. We reveal what these structures are in the right bottom row of Figure~\ref{fig:fig3}, using an annulus derived from our elliptic fits, as discussed in \S\ref{sec:less4}, but for the \texttt{M20Ma0.1} simulation. The bottom right panel reveals the density structures that are causing the $a > 1$ anisotropy. We see many, high-density, filamentary structures perpendicular to the magnetic field direction, consistent with orientation studies of the Gould Belt molecular clouds \citep{Planck2016a,Planck2016b}. Some parallel striations are still present, since the magnetic field is still dynamically important, but the perpendicular filaments dominate the morphology of the power spectrum.

\cite{Federrath2016} and \cite{Xu2019} independently suggested that the width of the high-density filaments,
\begin{equation} \label{eq:filament_width}
    \ell \propto L / \M^2,
\end{equation}
using two different models. \citet{Federrath2016} uses the turbulent sonic scale and \citet{Xu2019} use the Rankine-Hugoniot jump conditions to motivate Equation~(\ref{eq:filament_width}). In $k$-space, this means that $\kpar \sim \M^2$, and so once the filamentary structures become locked perpendicular to the $\vecB{B}$-field, increasing $\M$ grows the anisotropy along the $\kpar$ direction. This is because as the perpendicular filaments become thinner they contribute to more stretching along $\kpar$. Furthermore, since the number of filaments present in isothermal turbulence is in someway controlled by $\M$ \citep{Beattie2019c}, we expect that as $\M$ increases, more filaments develop, stretching the power spectrum further. Hence for observations of MCs with highly-orientated, high-density filaments, the magnetic field must be dynamically important, and at the same time, the turbulent motions must be significantly supersonic, i.e., $\M \gtrsim 4$.

\subsection{Summary}
We find that the anisotropy parameter $a$, defined in Equation~(\ref{eq:a}), has two dominant modes in the column density. The first mode, $a < 1$, is elongation along the $\kperp$ direction, which is due to striations that run parallel to the $\vecB{B}$-field, examined in \S\ref{sec:less4}. The $a < 1$ anisotropy persists across all $k$-scales, similar to the magnetic field, and indeed, seems to be largely controlled by the magnetic field. The second mode, $a > 1$, is elongation along the $\kpar$ direction in the highly-turbulent, strong $\vecB{B}$-field regime, discussed in \S\ref{sec:gtrM4}. This anisotropy coincides with the development of high-density filaments that form perpendicular to the magnetic field, and that are largely controlled by the turbulent Mach number (e.g., in their width). Thus, our results strongly suggest that MC observations revealing highly-orientated, high-density filaments are in a sub-Alfv\'enic, highly-supersonic regime ($\M > 4$). On large scales, the $a > 1$ anisotropy disappears, coinciding with large-scale isotropic, turbulent motions. More generally, for all simulations in the super-Alfv\'enic limit ($\Ma_0>1$), the density and flow structures are largely isotropic, regardless of $\M$, as discussed in \S\ref{sec:superAlf}.

It should be noted that our ensemble of simulations do not include any with gravitational dynamics, and the discussed anisotropic structures emerge purely in the presence of strongly-orientated magnetic fields and turbulent motions. Indeed, this suggests that these structures can be a product of magnetic fields and turbulence alone, however one might expect to observe more extreme and time-dependant anisotropic structures developing in clouds governed by gravito-magnetohydrodynamics. This is because gas that is accreted onto the high-density filamentary structures will cause gravitationally unstable filaments to collapse and create thin, orientated structures, further contributing to the stretching of the 2D power spectrum. We leave this investigation for future studies of the time-dependent anisotropy in collapsing molecular cloud systems.

\section{The 1D Power Spectra} \label{sec:1DPS}
Typical in spectral analysis is the reduction of the 2D power spectrum (or 3D power spectrum) into the 1D azimuthally-averaged power spectrum. However, the process of azimuthally averaging the power spectrum assumes that the turbulence is isotropic, i.e., $\vecB{k}$ is rotationally symmetric in $k$-space, hence the power spectrum only depends upon the modulus of $\vecB{k}$. For the anisotropic 2D power spectra discussed in \S\ref{sec:PS}, however, no such rotational symmetry exists in the sub-Alfv\'enic regime (at least in the LOS perpendicular to the magnetic field). For this reason, we decompose the 2D power spectra into $\kperp$ and $\kpar$ components, with respect to the orientation of the $\vecB{B}$-field. This has been done before in \cite{Bigot2008}, for example.

\begin{table}
\caption{Power-law scaling exponents for the perpendicular and parallel cascades in the zero-mean transformed column density.}
\centering
\begin{tabular}{lccc}
\hline
\hline
\multicolumn{1}{c}{Simulation} & $\alpha_{\perp}$ & $\alpha_{\parallel}$ & $\alpha_\parallel / \alpha_\perp$ \\
\multicolumn{1}{c}{ID} & $(\pm1\sigma)$ & $(\pm1\sigma)$ & $(\pm1\sigma)$  \\
\hline
\texttt{M2Ma10}\dotfill      & $-3.1 \pm 0.3$    & $-3.3 \pm 0.2$    & $1.1 \pm 0.1$ \\
\texttt{M2Ma2}\dotfill       & $-3.1 \pm 0.2$    & $-3.4 \pm 0.2$    & $1.1 \pm 0.1$ \\
\texttt{M2Ma1}\dotfill       & $-2.7 \pm 0.2$    & $-3.8 \pm 0.2$    & $1.4 \pm 0.1$ \\
\texttt{M2Ma0.5}\dotfill     & $-1.7 \pm 0.2$    & $-4.2 \pm 0.2$    & $2.5 \pm 0.3$ \\
\texttt{M2Ma0.1}\dotfill     & $-1.4 \pm 0.2$    & $-4.2 \pm 0.3$    & $3.0 \pm 0.5$ \\

\texttt{M4Ma10}\dotfill      & $-2.8 \pm 0.1$    & $-3.0 \pm 0.2$    & $1.1 \pm 0.1$ \\
\texttt{M4Ma2}\dotfill       & $-2.9 \pm 0.2$    & $-3.3 \pm 0.2$    & $1.1 \pm 0.1$ \\
\texttt{M4Ma1}\dotfill       & $-2.7 \pm 0.2$    & $-3.1 \pm 0.1$    & $1.2 \pm 0.1$ \\
\texttt{M4Ma0.5}\dotfill     & $-1.5 \pm 0.2$    & $-3.7 \pm 0.3$    & $2.5 \pm 0.3$ \\
\texttt{M4Ma0.1}\dotfill     & $-1.7 \pm 0.2$    & $-3.6 \pm 0.3$    & $2.1 \pm 0.3$ \\

\texttt{M10Ma10}\dotfill     & $-2.8 \pm 0.2$    & $-2.5 \pm 0.2$    & $0.9 \pm 0.1$ \\
\texttt{M10Ma2}\dotfill      & $-2.3 \pm 0.2$    & $-2.6 \pm 0.3$    & $1.1 \pm 0.2$ \\
\texttt{M10Ma1}\dotfill      & $-2.4 \pm 0.2$    & $-2.5 \pm 0.2$    & $1.0 \pm 0.1$ \\
\texttt{M10Ma0.5}\dotfill    & $-2.0 \pm 0.2$    & $-2.5 \pm 0.3$    & $1.3 \pm 0.2$ \\
\texttt{M10Ma0.1}\dotfill    & $-1.8 \pm 0.2$    & $-2.3 \pm 0.2$    & $1.3 \pm 0.2$ \\

\texttt{M20Ma10}\dotfill     & $-2.3 \pm 0.2$    & $-2.4 \pm 0.2$    & $1.0 \pm 0.1$ \\
\texttt{M20Ma2}\dotfill      & $-1.9 \pm 0.1$    & $-2.4 \pm 0.2$    & $1.3 \pm 0.1$ \\
\texttt{M20Ma1}\dotfill      & $-2.2 \pm 0.2$    & $-2.3 \pm 0.2$    & $1.0 \pm 0.1$ \\
\texttt{M20Ma0.5}\dotfill    & $-2.0 \pm 0.1$    & $-2.3 \pm 0.2$    & $1.2 \pm 0.1$ \\
\texttt{M20Ma0.1}\dotfill    & $-1.7 \pm 0.2$    & $-2.2 \pm 0.2$    & $1.3 \pm 0.2$ \\
\hline
\hline
\end{tabular} \\
\begin{tablenotes}
\item{\textit{\textbf{Notes:}} Column (1): the simulation ID. A full list of the simulation parameters for each ID is shown in Table \ref{tb:simtab}. Column (2): the perpendicular (to the magnetic field) power-law scaling exponent, $\mathscr{P}_\xi \propto \kperp^{\alpha_{\perp}}$ for the zero-mean transformed column density (Equation \ref{eq:meanfield}). Column (3): the same as column (2) but the parallel power-law scaling exponent, $\mathscr{P}_\xi \propto \kpar^{\alpha_{\parallel}}$. Column (4): the ratio between the two scaling exponents from columns (2) and (3).}
\end{tablenotes}
\label{tb:powerlaws}
\end{table}

\begin{figure*}
    \centering
    \includegraphics[width=\linewidth]{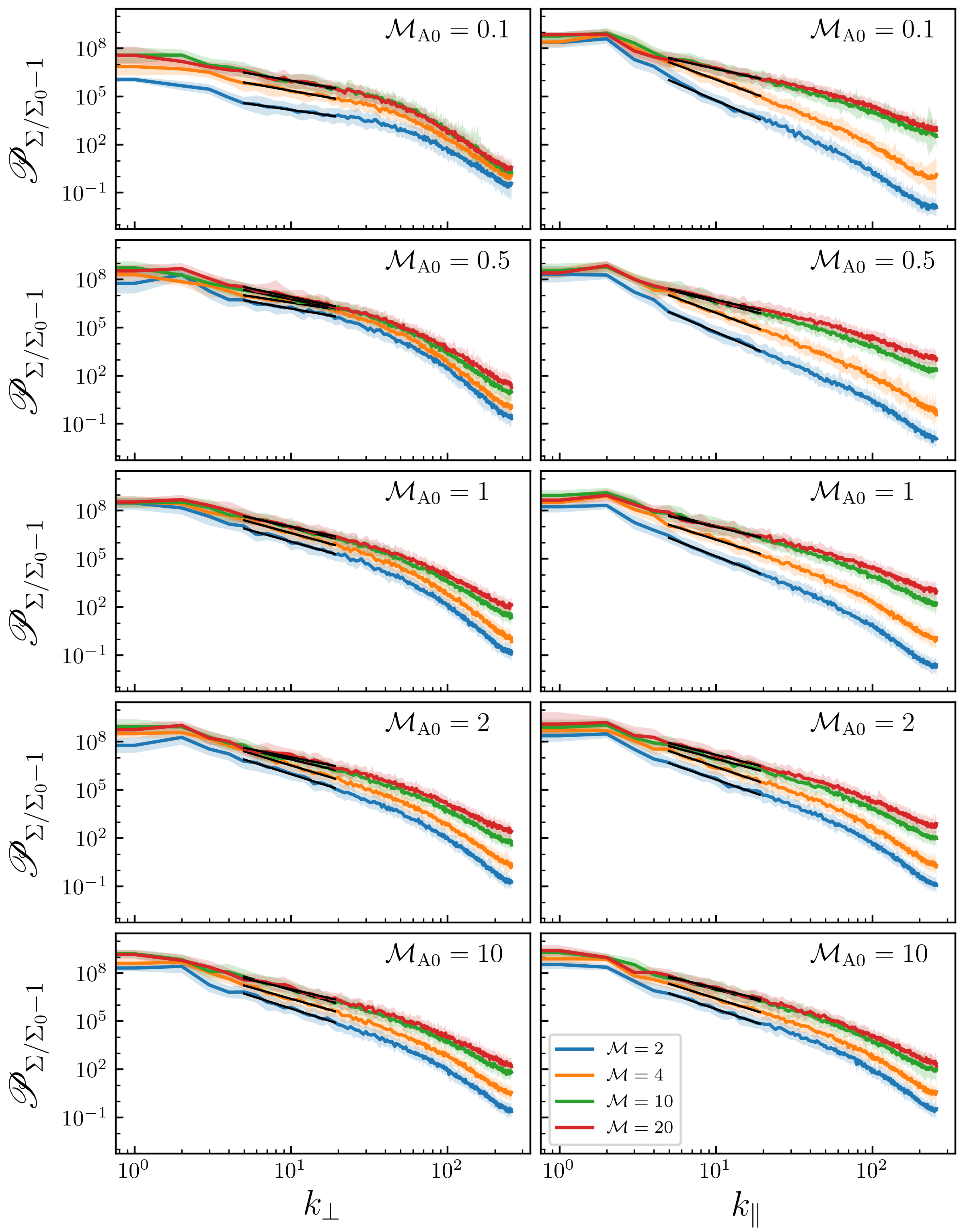}
    \caption{Slices of the 2D power spectra across the $\kperp$ (left panels) and $\kpar$ (right panels) directions. We fit slopes on power between $5 \leq k \leq 20$, illustrated in black, and tabulated in Table~\ref{tb:powerlaws}. The plots are coloured by $\M$, indicated in the legend of the bottom-right panel.}
    \label{fig:fig5}
\end{figure*}

\begin{figure*}
    \centering
    \includegraphics[width=\linewidth]{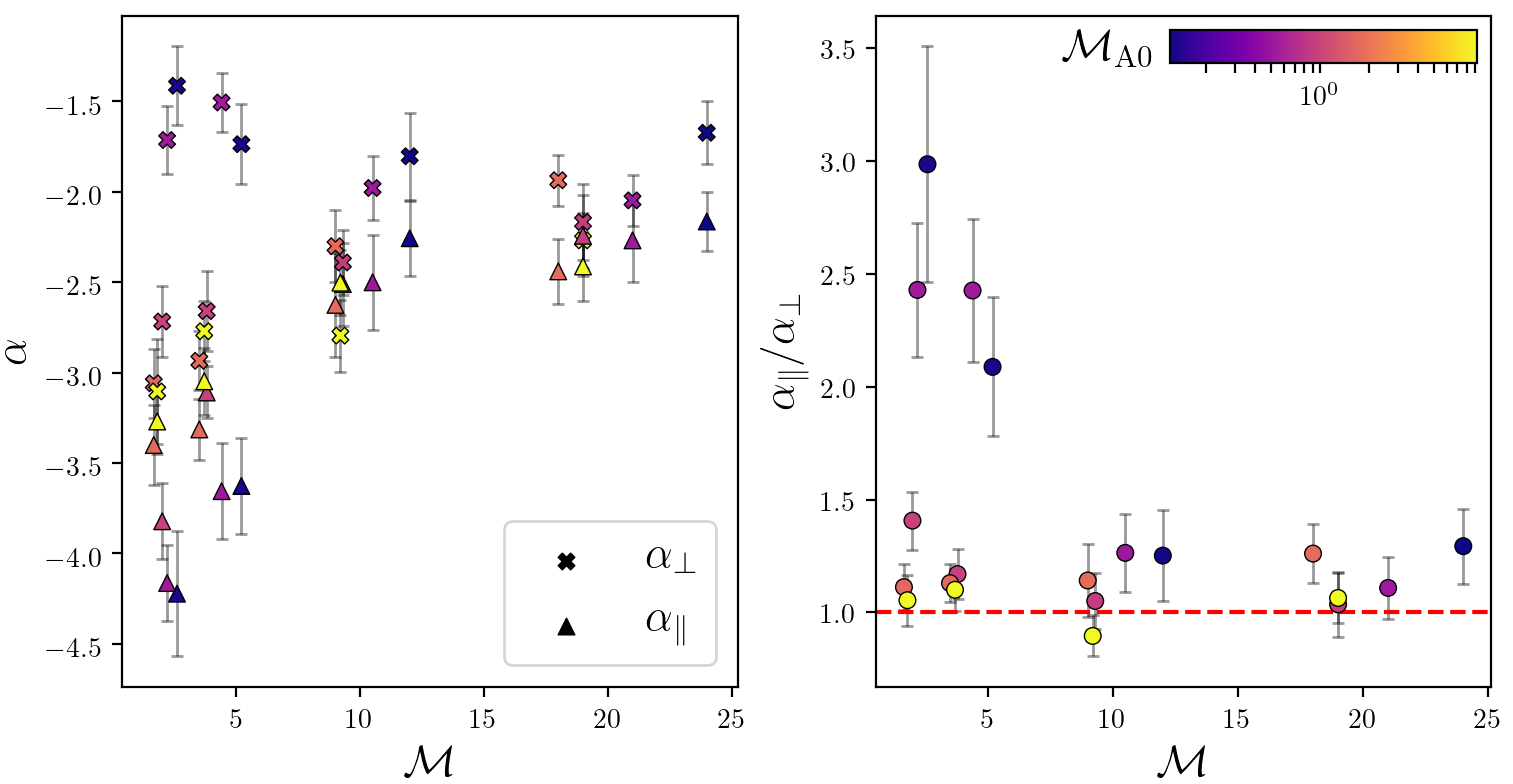}
    \caption{\textbf{Left:} power-law scaling exponents, $\alpha$, from the 1D sliced power spectra of the zero-mean transformed column density plotted as a function of turbulent Mach number. The crosses indicate the scaling exponent for the perpendicular component of the 1D power spectrum (Equation~\ref{eq:alphaperp}), and the triangles the parallel scaling exponent (Equation~\ref{eq:alphapar}). The markers are coloured by the Alfv\'en Mach number, where yellow is weak $\vecB{B}$-field, and dark-blue is strong $\vecB{B}$-field. \textbf{Right}: the same as the left panel but for the ratio between the parallel and perpendicular scaling exponents. We see the scaling exponents converging to the same value in the high-$\M$ limit, indicating the development of an isotropic 2D power spectrum.}
    \label{fig:fig6}
\end{figure*}

\subsection{Construction}
The simplest way to decompose the 2D power spectrum into 1D $\kperp$ and $\kpar$ components is to take slices through the 2D power spectrum. We slice at $\mathscr{P}_\xi(\kperp = 0, \kpar)$, for the $\kpar$ spectrum and  $\mathscr{P}_\xi(\kperp, \kpar=0)$, for the $\kperp$ spectrum. Next we fit power laws between wavenumbers $5 \leq k \leq 20$, which is a conservative estimate of where the scaling range might be found in the velocity structure for our simulations. We denote the power law parallel to the $\vecB{B}$-field
\begin{equation}\label{eq:alphapar}
    \mathscr{P}_\xi \propto \kpar^{\alpha_{\parallel}},
\end{equation}
and perpendicular to the $\vecB{B}$-field,
\begin{equation}\label{eq:alphaperp}
    \mathscr{P}_\xi \propto \kperp^{\alpha_{\perp}},
\end{equation}
where $\alpha_{\parallel}$ and $\alpha_{\perp}$ are the scaling exponents in the parallel and perpendicular cascades, respectively. We tabulate the exponents from the power-law fits, along with the ratio between them in Table \ref{tb:powerlaws}. Unlike the single-point statistics for the anisotropy parameters calculated in \S\ref{sec:aniso}, the power-law exponents computed here encode all the anisotropy information from length scales within the (approximate) cascades.

\subsection{1D Power Spectrum Results}
Figure~\ref{fig:fig5} shows that the slope of the parallel component of the power spectrum becomes shallower with increasing $\M$. This is because parallel to the $\vecB{B}$-field the densities do not feel the Lorentz force and are able to create higher-$k$ modes in the column density, corresponding to the development of density fluctuations and strong shocks \citep{Kim2005,Kowal2007}. Indeed, the strong shocks that develop from parallel compression along the $\vecB{B}$-field lines are what creates the high-density filaments perpendicular to the field, consistent with our results from \S\ref{sec:PS}. We plot the scaling exponents in Figure~\ref{fig:fig6} to better visualise the $\M$ and $\Ma_0$ dependencies. The triangle markers, corresponding to the parallel scaling exponents, have values of $\sim -4$ at low-$\M$, and are completely distinct from the perpendicular scaling exponents, shown with cross markers. However, with increasing $\M$, both of the scaling exponents become centralised around $\sim -2$, consistent with the reoccurring theme of isotropy in the high-$\M$ limit, discussed in the context of the 2D power spectrum in \S\ref{sec:superAlf}. 

Unlike the component of the column density parallel to the $\vecB{B}$-field, the perpendicular component feels very little of the turbulent motions. This is shown by tracing the $\alpha_{\perp}$ (cross markers) in the left panel of Figure~\ref{fig:fig6} with the same magnetic field strength (i.e., same colour) and observing how they stay relatively constant with changing $\M$. The spread of the scaling exponents seems to be determined by $\Ma_0$, where steeper values correspond to weaker fields, similar to the average anisotropic parameters discussed in \S\ref{sec:aniso}. This is consistent with the ``magnetic cushioning" principle described in \citet{Molina2012}, whereby the $\vecB{B}$-field suppresses fluctuations in the density field, resulting in a steeper slope in the power spectrum. 

In the right-hand panel of Figure~\ref{fig:fig6} we show the ratio between the parallel and perpendicular scaling exponents as a function of $\M$. In the low-$\Ma_0$, low-$\M$ power spectra, to the left of the panel, $\alpha_{\parallel}$ is $\sim 2.5$, steeper than $\alpha_{\perp}$. This means that most of the high-$k$ density fluctuations are transverse to the field lines, which could indeed be associated with the striations discussed in \S\ref{sec:subAlf}. However, as $\M$ increases, unlike the 2D power spectrum, there is no distinct signature of the high-density filaments.

\begin{figure*}
    \centering
    \includegraphics[width=\textwidth]{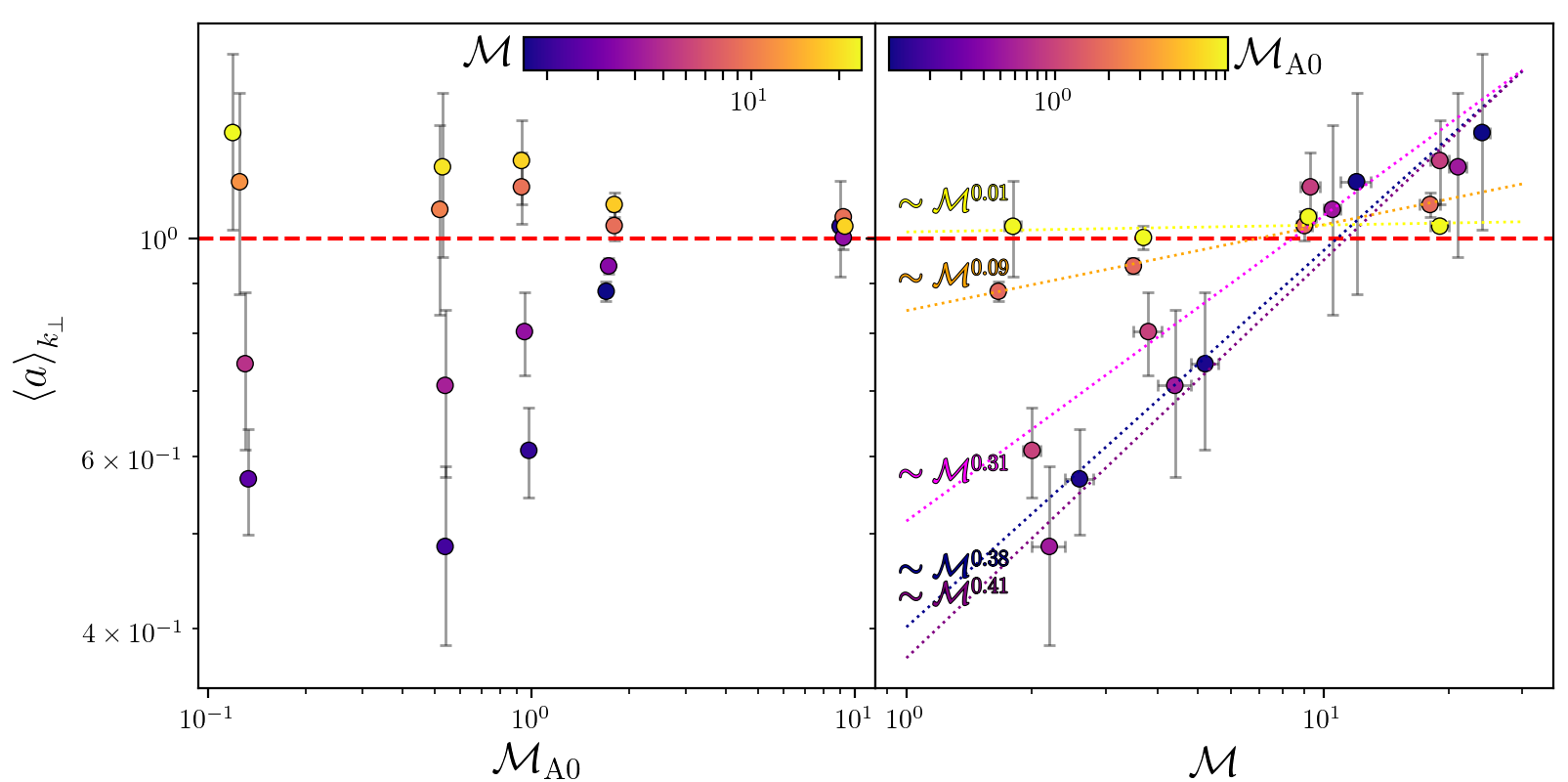}
    \caption{\textbf{Left Panel:} the average anisotropy, $\Exp{a}_{\kperp}$ (Equation~\ref{eq:averAniso}), in the zero-mean transformed column density (Equation~\ref{eq:meanfield}) as a function of $\Ma_0$, coloured by $\M$. The red, dashed line indicates perfect isotropy at $\Exp{a}_{\kperp} = 1$. Above the isotropic boundary are the clouds with high-density filaments dominating the morphology of the power spectrum, which we examine in \S\ref{sec:gtrM4}, and clouds below are dominated by magnetosonic striations, which we discuss in \S\ref{sec:less4}. \textbf{Right Panel:} the same as the left panel, but as a function of $\M$ and coloured by $\Ma_0$. We fit power laws to data with the same $\Ma_0$. The fits are indicated with the dotted lines, and labelled by the power law, $\Exp{a}_{\kperp} \sim \M^\beta$. Decreasing $\Ma_0$ steepens the power law, increasing the overall anisotropy, but whether or not the average anisotropy is above or below unity is controlled by $\M$.}
    \label{fig:fig4}
\end{figure*}

\section{The Average Anisotropy in the Column Density} \label{sec:aniso}
In \S \ref{sec:PS} we qualitatively described the anisotropy present in the column densities as a function of scale in the 2D power spectra. Next in \S \ref{sec:1DPS} we explored how the 1D power spectra perpendicular and parallel to the $\vecB{B}$-field encode some (but not all) information about the anisotropic structures found in \S \ref{sec:PS}. Now we summarise the anisotropy for each model using a single, scale-averaged anisotropy parameter. We define our scale-averaged anisotropy parameter as
\begin{equation}\label{eq:averAniso}
    \Exp{a}_{\kperp} =  \frac{1}{\kperp_{\text{max}}}\displaystyle \int\displaylimits^{\kperp_{\text{max}}}_0 \d{\kperp} \,  a(\kperp),
\end{equation}
where we integrate across all anisotropy parameters for all available $\kperp$-scales, and divide by the maximum $\kperp$, $\kperp_{\text{max}}$, which is $\kperp$ at the last closed contour in the 2D power spectrum. We show examples of the $a(\kperp)$ curves for the two extreme anisotropic simulations ($\M=2$ and $\M=20$) in Appendix \ref{sec:scaleAnisoApp}. We do this for each power spectrum, creating a single number that represents the average anisotropy of the 2D power spectrum, tabulated in column (9) of Table \ref{tb:simtab}, shown with $1\sigma$ uncertainties defined as
\begin{equation} \label{eq:aaver}
     \sigma_{\Exp{a}_{\kperp}} = \sqrt{ \Exp{  a ^2}_{\kperp} - \Exp{a}_{\kperp}^2 }.
\end{equation}
This is similar to what has been done in \cite{Appleby2018}, where they use Minkowski tensors to approximate the anisotropy of dark matter fields. Our anisotropy parameter is analogous to taking the average of the ratio between the two eigenvalues of the $(W_2^{1,1})_{ij}$ tensor (equation 15 in \citealt{Appleby2018}) over all excursion sets on the 2D power spectrum. 

\subsection{$\Exp{a}_{\kperp}$ Results}

\begin{figure}
    \centering
    \includegraphics[width=\linewidth]{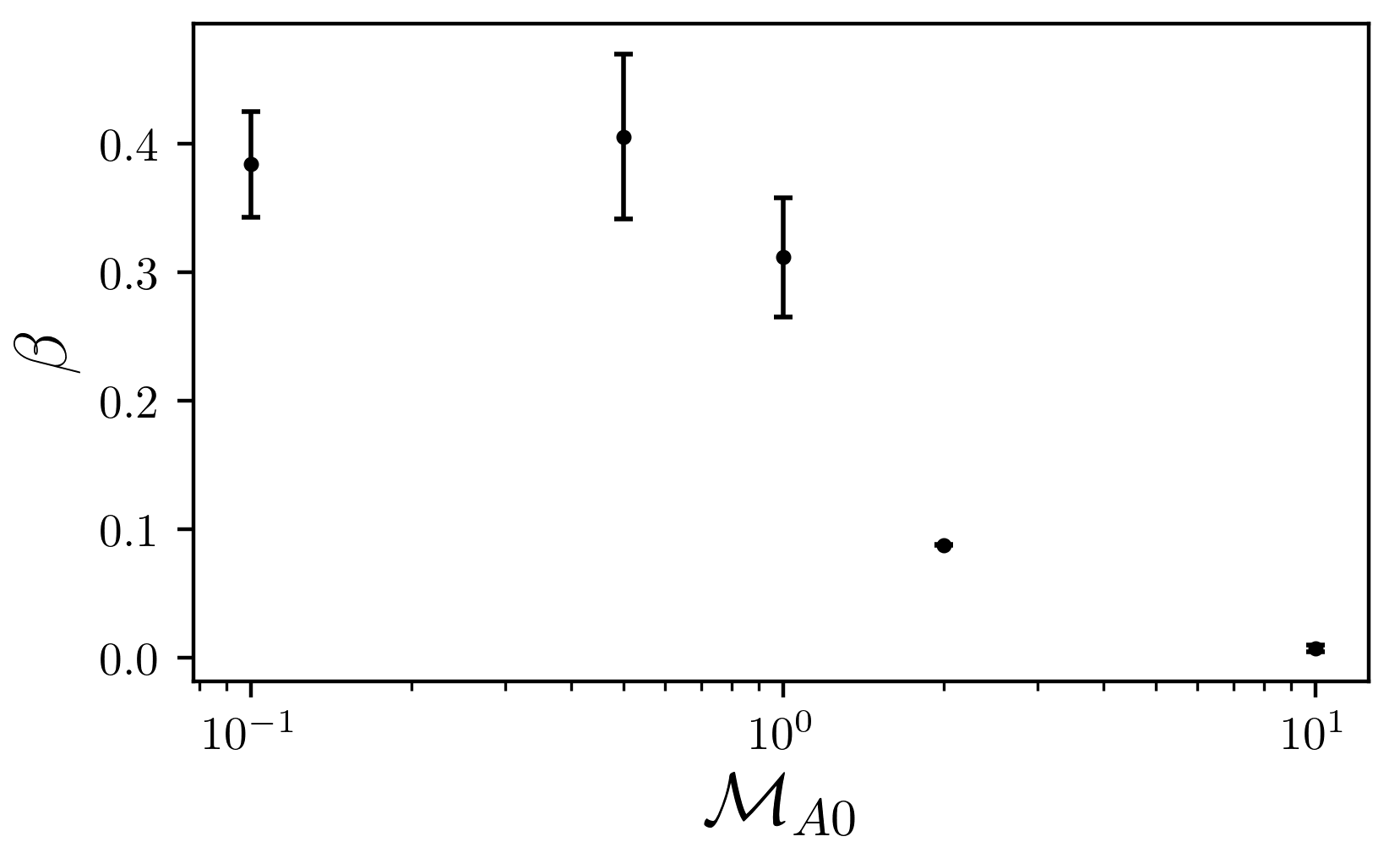}
    \caption{The power-law exponent, $\beta$, for the $\Exp{a}_{\kperp}$ -- $\M$ relation, shown in Equation \ref{eq:a(M)} and illustrated in Figure \ref{fig:fig4}, as a function of $\Ma_0$. We see a smooth transition across the values of $\Ma_0$ that is bounded above by $\beta \sim 0.4$ and below by $\beta \sim 0$.}
    \label{fig:beta}
\end{figure}

We show the average anisotropy parameter $\Exp{a}_{\kperp}$, defined in Equation~(\ref{eq:aaver}), as a function of $\Ma_0$ and $\M$ in Figure \ref{fig:fig4}. In the left plot we see the average anisotropy as a function of $\Ma_0$, coloured by $\M$. There are two key features in this plot. First, as $\Ma_0$ increases, the average anisotropy disappears (i.e., $\Exp{a}_{\kperp}$ approaches unity). Second, for a fixed $\Ma_0$, the spread is controlled by $\M$. To understand this relationship we show the average anisotropy as a function of $\M$ in the right-hand panel of Figure~\ref{fig:fig4}. For a fixed $\Ma_0$ we find that there exits a power law in $\M$,
\begin{equation}\label{eq:a(M)}
    \Exp{a}_{\kperp} \propto \M^\beta.
\end{equation}
The power laws may not strictly hold for the very high- and low-$\M$ limits (e.g., there may be an anisotropic saturation), but they provide a reasonable approximation for the intermediate regime covered by the simulations. We show the fits with the dotted lines, coloured by $\Ma_0$ in Figure~\ref{fig:fig4}. At $\Ma_0 \approx 10$ (shown in yellow) we find that the average anisotropy is approximately constant in $\M$, $\beta \sim 0$, shown to the right of Figure \ref{fig:beta}, hence in the weak $\vecB{B}$-field limit, no anisotropy exists for all $\M$, consistent with our previous qualitative investigations in \S\ref{sec:PS}. However, in the strong- $\vecB{B}$-field regime we see the slope of the power laws increase until reaching a saturation at $\beta \sim 0.4$ when $\Ma_0 \sim 0.1 - 0.5$, shown to the left of Figure \ref{fig:beta}. As we move along the $\M$-axis, we capture the transition from the striation-dominated regime, $\Exp{a}_{\kperp}<1$, to the high-density, filament-dominated regime, $\Exp{a}_{\kperp}>1$. This shows that the type of anisotropy, $a > 1$ or $a < 1$, is controlled by $\M$, but when (at what $\M$) the transition occurs is set by $\Ma_0$, which is a key result from this study. Indeed, this means that one may be able to use the anisotropy in the 2D power spectrum to determine important cloud parameters, such as the mean component of the magnetic field or the sonic and Alfv\'enic Mach numbers, which will be explored in a future study. Next we move onto another measure of the anisotropy, the perpendicular and parallel 1D power spectra of the column density.

\section{Summary and Key Findings}\label{sec:conclusion}
In this study we use 51 time-realisations from 20 high-resolution, turbulent magnetohydrodynamical (MHD) simulations with mean-field Alfv\'en Mach numbers, $0.1 \leq \Ma_0 \leq 10$, and turbulent Mach numbers, $2 \leq \M \leq 20$ (simulation parameters are shown in Table \ref{tb:simtab}). We create two-dimensional (2D) projections (column densities) perpendicular to the threaded magnetic fields, shown in Figure~\ref{fig:fig1}. We construct 2D power spectra for each of the zero-mean transformed column densities (Equation~\ref{eq:meanfield}), and average them over five turbulent crossing times to show how the average equi-power structures are organised in $k$-space, shown in Figure~\ref{fig:fig2}. We calculate the anisotropy for each simulation by fitting ellipses to the contours of the 2D power spectra and calculate the ratio, $a$, between the minor and major principle axis of the fitted ellipses. We reveal two distinct anisotropies, $a < 1$, corresponding to magnetosonic striations, and $a > 1$, corresponding to high-density filaments, distinguished in Figure~\ref{fig:fig3}, and discussed in detail in \S\ref{sec:PS}. Next we create 1D slices of the 2D power spectra, shown in Figure~\ref{fig:fig5}, and discuss how the perpendicular and parallel scaling exponents in the slices reveal some similarities to our 2D spectral analysis in \S\ref{sec:PS} and depend upon the magnetic field and turbulence parameters. In \S\ref{sec:aniso} we average the anisotropy parameter, $a$, across all spatial scales in each simulation and show how it depends upon $\M$ and $\Ma_0$ in Figure~\ref{fig:fig4}. In the following, we list the keys results from this study.

\begin{itemize}
    \item The 2D power spectrum of the column density (Figure~\ref{fig:fig2}) reveals two distinct types of anisotropy, $a$, in the sub-Alfv\'enic MHD turbulence regime, neither of which are present in the super-Alfv\'enic regime. Both types of anisotropies exhibit scale-dependency, discussed in detail in \S\ref{sec:less4} and \S\ref{sec:gtrM4}.\\[1em]
    
    \item The column density of sub-Alfv\'enic, $\M \lesssim 4$ turbulence is dominated by magnetosonic striations running parallel to the magnetic field (top panels in Figure~\ref{fig:fig3}). This results in significant stretching of the 2D power spectrum along the $\kperp$ axis, which is defined as an $a < 1$ anisotropy using our anisotropy definition (\S\ref{sec:less4}), and shallows the slope of the perpendicular cascade in the 1D perpendicular power spectrum (\S\ref{sec:1DPS}). The anisotropy from the striations becomes stronger on large (box) scales, since the long structures, parallel to the $\vecB{B}$-field, span across the full box-length. The presence of striations in the observed column densities of molecular clouds may therefore mean that the cloud has a dynamically important magnetic field. \\[1em]  
    
    \item For the sub-Alfv\'enic, $\M \gtrsim 4$ simulations, high-density filaments, perpendicular to the magnetic field stretch the 2D power spectrum of the column density along the $\kpar$ axis (bottom panels of Figure~\ref{fig:fig3}), resulting in an $a > 1$ anisotropy, distinct from the $a < 1$ anisotropy in $\M \lesssim 4$ turbulence (\S\ref{sec:gtrM4}). For this regime, the anisotropy is only present on scales smaller than the box-length. At the box-length the turbulence returns to the isotropic regime. Indeed, molecular cloud observations that reveal highly-orientated, high-density filaments are therefore likely to be in the sub-Alfv\'enic, highly-supersonic ($\M \gtrsim 4$) dynamical regime. \\[1em]
    
    \item The anisotropy parameter averaged over all spatial scales, $\Exp{a}_{\kperp}$, approximately follows a power law for fixed magnetic field strength in $\M$, $\Exp{a}_{\kperp} \propto \M^\beta$ (see \S\ref{sec:aniso} and Figure~\ref{fig:fig4}). The value of $\beta$ changes from $\sim 0 - 0.4$ (see Figure~\ref{fig:beta}) for $\Ma_0 \approx 10 - 0.1$, respectively. This means that $\Ma_0$ largely controls the degree of anisotropy, but it is $\M$ that controls the transition between $a < 1$ and $a > 1$, i.e., the stretching of the 2D power spectrum either along the $\kperp$ or $\kpar$ axis, respectively. 
\end{itemize}

The quantification of anisotropies in the column density may be useful for constraining cloud parameters such as $\Ma_0$ and $\M$ in follow-up studies.

\section*{Acknowledgements}\label{sec:acknowledgments}
We~thank the anonymous reviewer for the useful suggestions that improved this study. J.~R.~B.~thanks Jonah Hansen for helpful editorial comments. C.~F.~acknowledges funding provided by the Australian Research Council (Discovery Project DP170100603, and Future Fellowship FT180100495), and the Australia-Germany Joint Research Cooperation Scheme (UA-DAAD). We further acknowledge high-performance computing resources provided by the Leibniz Rechenzentrum and the Gauss Centre for Supercomputing (grants~pr32lo, pr48pi and GCS Large-scale project~10391), the Partnership for Advanced Computing in Europe (PRACE grant pr89mu), the Australian National Computational Infrastructure (grant~ek9), and the Pawsey Supercomputing Centre with funding from the Australian Government and the Government of Western Australia, in the framework of the National Computational Merit Allocation Scheme and the ANU Merit Allocation Scheme. The simulation software \textsc{flash} was in part developed by the DOE-supported Flash Centre for Computational Science at the University of Chicago.

\bibliographystyle{mnras.bst}
\bibliography{new.bib}

\appendix

\section{Inversion into Real-space}\label{sec:invert}
We can access anisotropic features from the 2D power spectrum and put them back into real-space by convolving a top hat mask with the same elliptic properties of the contours to the 2D Fourier transform of the density field, and then inverting it. First we define the mask,
\begin{equation}
    \mathcal{E}_{\sigma} = \mathcal{N}(\sigma) * \mathcal{E}(e_{\parallel,1},e_{\perp,1},e_{\parallel,2},e_{\perp,2}),
\end{equation}
where, $\mathcal{E}$, is the annular region defined between two ellipses with principle axes ($e_{\parallel,1},e_{\perp,1}$) and ($e_{\parallel,2},e_{\perp,2}$), $\mathcal{N}$ is a Gaussian smoothing function, which smooths the annulus with kernel size $\sigma$ and $*$ is the convolution operator. We smooth the annulus to avoid the Gibbs phenomenon at the discontinuous edge of the annulus, which will otherwise be present after taking the inverse Fourier transform. We choose $\sigma = 2\,\d{x}$, where $\d{x}$ is the size of a grid-cell of the $\mathcal{E}$ mask. We choose this because it sufficiently smooths the discontinuous edge of the mask, whilst minimising the amount of overflow into neighbouring grid-cells (which will correspond to neighbouring $k$-modes). Next we take the convolution with the 2D Fourier transform, shown in Equation \ref{eq:FT}, of the column density, $\hat{\xi}$, where we remind the reader, $\xi = \Sigma / \Sigma_0 - 1$ is the zero-mean transformed column density, and invert the convolved Fourier image of the column density into real-space, i.e.,
\begin{align}
    \xi_{\mathcal{E}} &= \mathscr{F}^{-1}\left[\hat{\xi} * \mathcal{E}_{\sigma} \right] \\
    &= \displaystyle\sum^{N-1}_{k_x = 0} \sum^{N-1}_{k_y = 0} \left[\hat{\xi} * \mathcal{E}_{\sigma} \right] \exp\left\{ i\frac{\vecB{\ell}_{xy} \cdot \vecB{k}_{xy} }{N}  \right\}. 
\end{align}
Since the inverse Fourier transform is a complex-valued function we expect there to be a real and imaginary component to $\xi_{\mathcal{E}}$. We take just the real structures and plot them in the third row of Figure \ref{fig:fig3}. In Figure \ref{fig:fig3} we use the annulus $18 \leq \kperp \leq 58 \cup 10 \leq \kpar \leq 27$ and $18 \leq \kperp \leq 62 \cup 17 \leq \kpar \leq 84$, for the \texttt{M2Ma0.1} and \texttt{M20Ma0.1} simulation, respectively. These regions are extracted directly from the elliptic fits, shown in Figure \ref{fig:fig2}. This method reveals what kind of structures in the column density maps are responsible for the anisotropic features in the 2D power spectra.

\begin{figure}
    \centering
    \includegraphics[width=\linewidth]{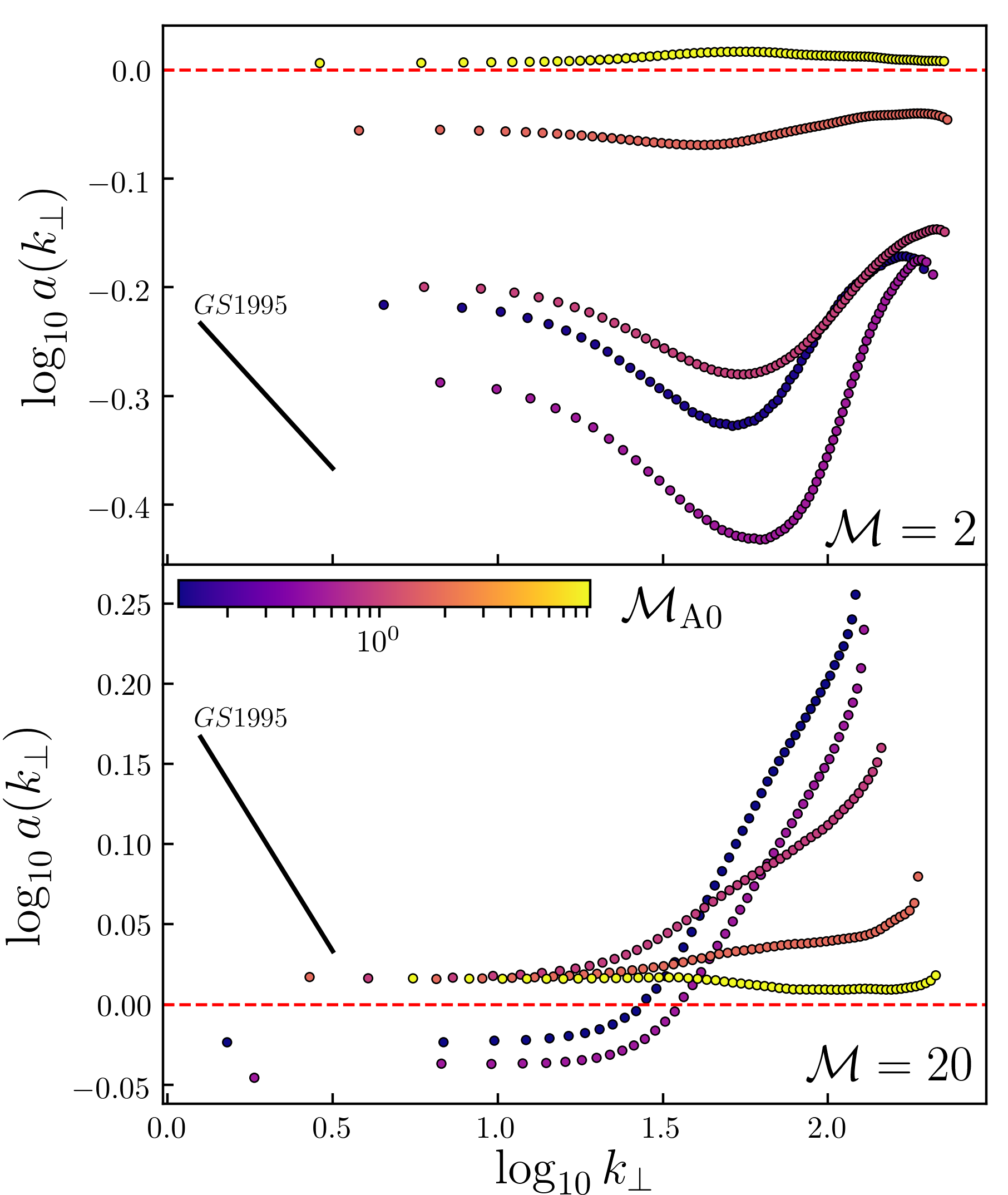}
    \caption{The anisotropy parameter, $a$, shown in Equation \ref{eq:a}, as a function of wave vector perpendicular to the guide field, $\kperp$ for the two extreme anisotropic simulations examined in \S \ref{sec:PS}, coloured by mean-field Alv\'en Mach number, $\Ma_0$. The \protect\cite{Goldreich1995}(GS1995) scaling, Equation \ref{eq:GSscaling}, is shown in black, which is the anisotropy predicted in the energy spectrum for incompressible MHD turbulence. We find that the spectra for the column density especially in the $\M = 20$ simulation do not follow the GS1995 scaling.}
    \label{fig:AppendixScaleDepend}
\end{figure}

\section{Anisotropy in the column density as a function of $\kperp$}\label{sec:scaleAnisoApp}

We show the anisotropy, $a$, as a function of scale, $\kperp$ in Figure \ref{fig:AppendixScaleDepend} for the $\M=2$ (top plot) and $\M=20$ (bottom plot) simulations, varying from $\Ma_0 = 0.1 - 10$. These are illustrative of the curves that we use to average over in \S\ref{sec:aniso} to create the single spatially-averaged statistic, $\Exp{a}_{\kperp}$, for each simulation. The \cite{Goldreich1995} (GS1995) scaling, shown in Equation \ref{eq:GSscaling}, is illustrated in black, at the left of each of the plots. 

The GS1995 scaling is steeper than the observed decline in the $\M=2$ simulations and in the wrong direction for the $\M=20$ simulations. This result is not entirely surprising, since the critical balance in GS1995 is concerned with the energy spectrum of incompressible MHD turbulence, and in this study we are concerned with the power spectrum of the column density in highly-compressible MHD turbulence -- two very different quantities. 

\section{Anisotropy in the Column Density as a function of LOS projection Angles}\label{sec:AppProj}

\begin{figure*}
    \centering
    \includegraphics[width=\linewidth]{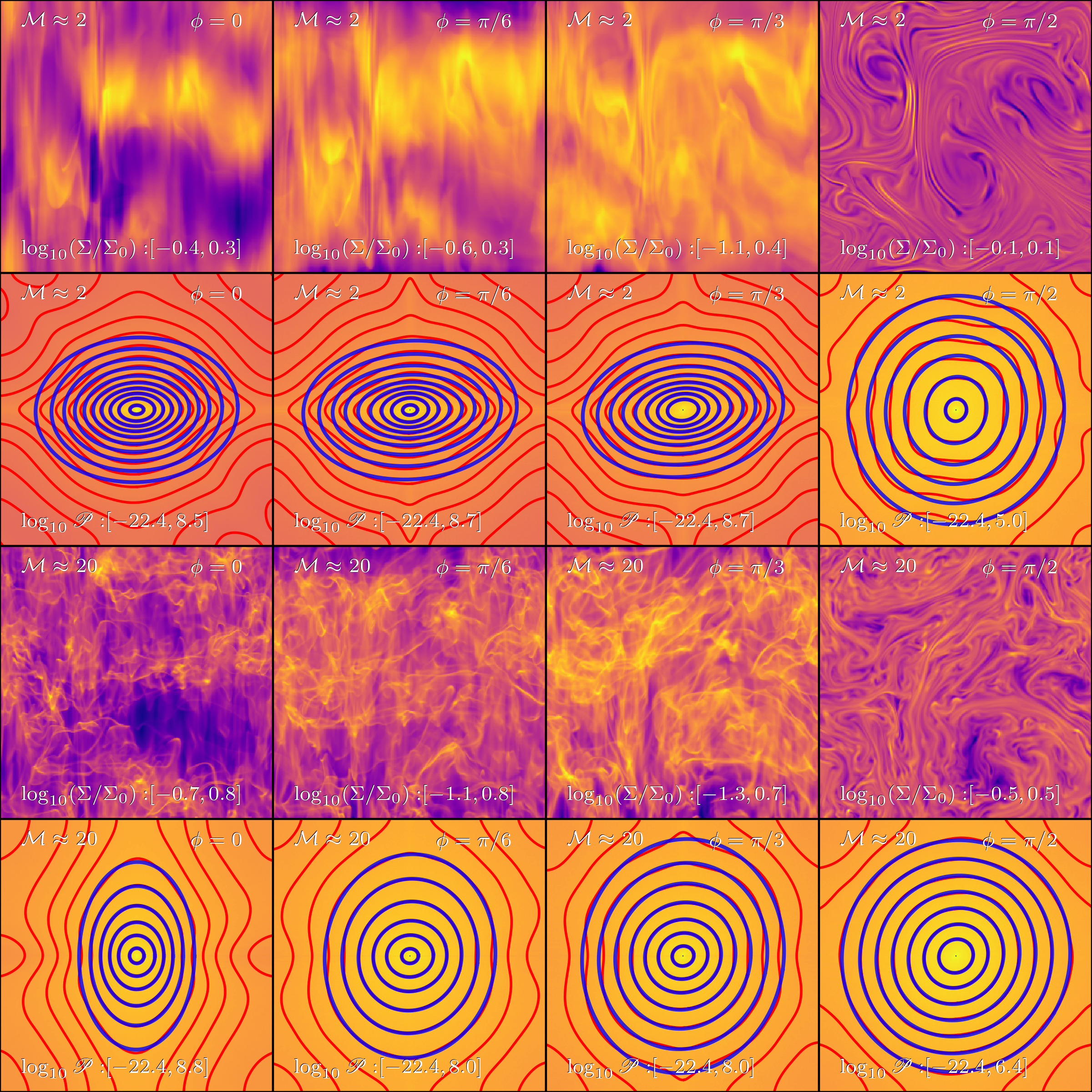}
    \caption{The column density maps and 2D power spectra for the \texttt{M2Ma0.1} (top two rows) and \texttt{M20Ma0.1} (bottom two rows) simulations shown at line-of-sight (LOS) projection angles $\phi = 0,\, \pi/6,\, \pi/3$ and $\pi/2$, from left to right, respectively. $\phi = 0$ corresponds to the LOS perpendicular to the magnetic guide field and $\phi = \pi/2$ the LOS parallel to the guide field. As the LOS moves towards the parallel direction we find that the anisotropies discussed in the main text disappear, revealing approximately isotropic behaviour around the magnetic guide field in the column density.}
    \label{fig:AppendixMaps}
\end{figure*}

\begin{figure}
    \centering
    \includegraphics[width=\linewidth]{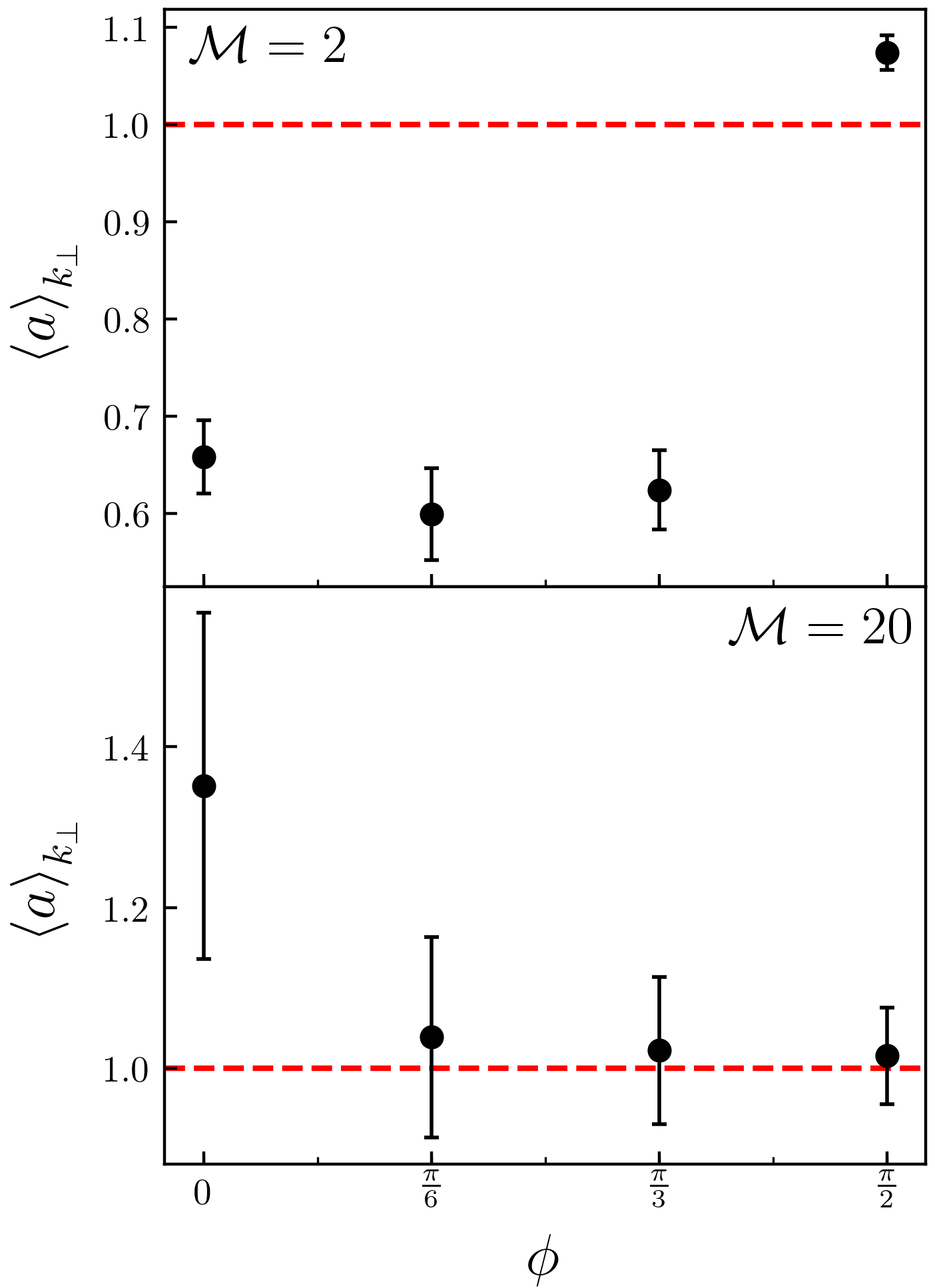}
    \caption{The average anisotropy, $\Exp{a}_{\kperp}$, as described in \S\ref{sec:aniso}, as a function of line-of-sight (LOS) viewing angle, $\phi$, where $\phi = 0$ and $\phi = \pi/2$ correspond to the LOS viewing angle perpendicular and parallel to the magnetic guide field, respectively. The red line indicates perfect isotropy. The top panel is for the \texttt{M2Ma0.1} simulation and the bottom for \texttt{M20Ma0.1}. We show examples of the column density maps for each $\phi$ in Figure \ref{fig:AppendixMaps}.}
    \label{fig:AppendixAngles}
\end{figure}

We explore the anisotropy as a function of line-of-sight (LOS) viewing angle, $\phi$, with respect to the perpendicular viewing angle to the magnetic guild field, $\vecB{B} = B_z \hat{z}$, as explored in the main text. We call this $\phi = 0$. We focus our analysis on the most anisotropic simulations, \texttt{M2Ma0.1} and \texttt{M20Ma0.1}. We show the projections and 2D power spectra for $\phi = 0,\, \pi/6,\, \pi/3$ and $\pi/2$, in Figure \ref{fig:AppendixMaps}, following the methods from \S\ref{sec:PS}, where $\phi = \pi/2$ is the projection that runs along the guide field.\\ 
Figure \ref{fig:AppendixMaps} reveals that the power spectrum becomes more isotropic as our viewing angle moves between the perpendicular and parallel viewing angles with respect to the guide field. We calculate the average anisotropy factor, as we did in \S\ref{sec:aniso} and plot it as a function of $\phi$ in Figure \ref{fig:AppendixAngles}. We find that, for the highly-supersonic simulation, \texttt{M20Ma0.1} there is a smooth transition towards isotropy as the viewing angle coincides with the direction of the magnetic field. However, for the \texttt{M2Ma0.1} there is an abrupt change into the isotropic regime only when the viewing angle is directly aligned with the magnetic field. This means that the anisotropy caused by the striations is more robust to changes in th viewing angle than the anisotropy caused by the perpendicular filaments.

\section{3D density visualisations}
\begin{figure*}
    \centering
    \includegraphics[width=0.93\linewidth]{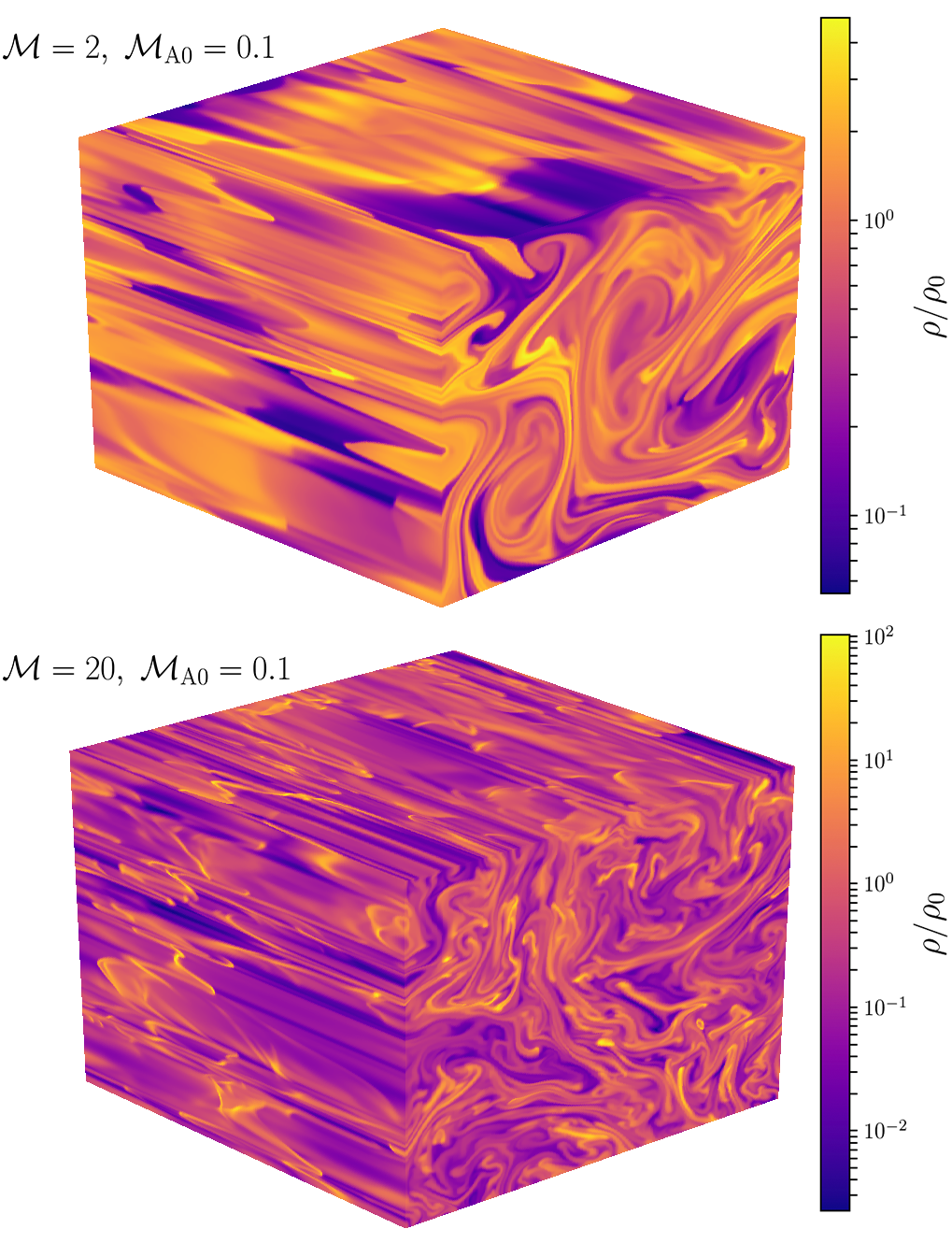}
    \caption{The gas density shown in 3D, for the two simulations, \texttt{M2Ma0.1} and \texttt{M20Ma0.1}, exhibiting extreme anisotropy along the direction of the magnetic field lines. Both 3D densities reveal that striations are most likely long, thin, sheets that twist around the magnetic field lines due to the turbulent motions. Furthermore, in the \texttt{M20Ma0.1} simulation we see high-density filaments forming parallel to the magnetic fields.}
    \label{fig:Appendix1}
\end{figure*}

\label{lastpage}

\end{document}